\documentclass[journal]{IEEEtran}
\usepackage{cite}
\usepackage{amsmath,amssymb,amsfonts}
\usepackage{algorithmic}
\usepackage{graphicx}
\usepackage{float}
\usepackage{textcomp}
\def\BibTeX{{\rm B\kern-.05em{\sc i\kern-.025em b}\kern-.08emT\kern-.1667em\lower.7ex\hbox{E}\kern-.125emX}}

\usepackage[ruled,norelsize]{algorithm2e}

\usepackage{url}
\newtheorem{thm}{Theorem}
\newtheorem{mydefn}{Definition}

\newtheorem{lem}{Lemma}
\newtheorem{prob}{Problem}

\newtheorem{assm}{Assumption}
\newtheorem{reform}{Reformulation}

\newcommand{\R}{\mathbb{R}}
\newcommand{\N}{\mathbb{N}}
\renewcommand{\P}{\mathbb{P}}
\newcommand{\E}{\mathbb{E}}

\newcommand{\pr}[1]{\P\!\left(#1\right)}
\newcommand{\ex}[1]{\E\!\left[#1\right]}
\newcommand{\tr}[1]{\mathrm{tr}\!\left(#1\right)}
\newcommand{\var}[1]{\mathrm{Var}\!\left(#1\right)}
\newcommand{\std}[1]{\mathrm{Std}\!\left(#1\right)}
\newcommand{\cov}[2]{\mathrm{Cov}\!\left(#1,#2\right)}
\newcommand{\bvec}[1]{\vec{\boldsymbol{#1}}}
\newcommand{\Nt}[2]{\N_{[#1,#2]}}

\makeatletter
\let\NAT@parse\undefined
\makeatother
\usepackage{hyperref}
\hypersetup{
    colorlinks=true,      
    linkcolor=black,
    citecolor=black,
    filecolor=black,
    urlcolor=black     
    }

\title{Chance Constrained Stochastic Optimal Control for Arbitrarily Disturbed LTI Systems Via the One-Sided Vysochanskij–Petunin Inequality} 
\author{Shawn Priore, \IEEEmembership{Student Member, IEEE}, and Meeko Oishi, \IEEEmembership{Senior Member, IEEE}
\thanks{This material is based upon work supported by the National Science Foundation under NSF Grant Number CMMI-2105631. (Corresponding author: Shawn Priore)} 
\thanks{Shawn Priore, and Meeko Oishi are with the Department of Electrical and Computer Engineering, University of New Mexico, Albuquerque, NM 87131 (e-mail: \texttt{shawn.a.priore@gmail.com, oishi@unm.edu}).
    }
}

\begin{document}
\maketitle

\begin{abstract}
    While many techniques have been developed for chance constrained stochastic optimal control with Gaussian disturbance processes, far less is known about computationally efficient methods to handle non-Gaussian processes. In this paper, we develop a method for solving chance constrained stochastic optimal control problems for linear time-invariant systems with general additive disturbances with finite moments and unimodal chance constraints. We propose an open-loop control scheme for multi-vehicle planning, with both target sets and collision avoidance constraints. Our method relies on the one-sided Vysochanskij–Petunin inequality, a tool from statistics used to bound tail probabilities of unimodal random variables. Using the one-sided Vysochanskij–Petunin inequality, we reformulate each chance constraint in terms of the expectation and standard deviation. While the reformulated bounds are conservative with respect to the original bounds, they have a simple and closed form, and are amenable to difference of convex optimization techniques. We demonstrate our approach on a multi-satellite rendezvous problem. 
\end{abstract}


\section{Introduction} \label{sec:intro}

Autonomous systems that are high risk, expensive, or safety critical require assurances they will not enter unsafe conditions that may lead to costly damage to property or loss of life. Satellite constellations and self-driving cars are just two examples where failure can be prohibitively expensive. Stochasticity, such as that due to modeling errors, external forces, or incomplete knowledge of the environment, complicates efforts to provide formal assurances in autonomous systems. Probabilistic assurances, while not as strong as those based in robust approaches that presume a worst-case scenario, allow for assurances tailored to a desired level of confidence or risk. However, although many stochastic effects are non-Gaussian (such as heavy tail phenomena in relative satellite dynamics), few methods exist that can accommodate non-Gaussian stochastic processes within a stochastic optimal control framework. 

One of the primary challenges associated with stochastic optimal control with non-Gaussian processes is the lack of analytic expressions for the cumulative distribution function (CDF) of the state as it evolves over time. This deficiency is relevant for the evaluation of chance constraint probabilities, and typically requires high dimensional and often intractable integration. Characteristic function based approaches utilize closed form expressions in the Fourier domain to approximate the CDF with numerical Fourier inversions, but are limited to evaluation of chance constraints for convex sets \cite{Idan2019, vinod2019piecewise, Sivaramakrishnan2021TAC}. Simulation based approaches \cite{ono2008iterative, calafiore2006scenario, campi2011sampling} bypass the need for integration, but are reliant upon on the quality and size of the samples. Further, in practice, these approaches may be additionally limited by computational memory, necessary for large samples, as well as the need to sample the distribution. Sample reduction methods \cite{campi2011sampling, care2014fast, Campi2018TAC} decrease computational burden, by focusing on scenario characteristics and comparing them with previous solutions. However, the characteristic function approach requires numerical approximations of the CDF or its inverse \cite{Sivaramakrishnan2021TAC}, and the sampling approaches produce confidence bounds on chance constraint satisfaction \cite{campi2011sampling}, both of which weaken guarantees.

In contrast, methods that employ concentration inequalities provide almost surely assurances of chance constraint satisfaction through over-approximations. Chebyshev's inequality \cite{Boucheron2013} and Cantelli's inequality \cite{Boucheron2013} have been used to develop chance constraint reformulations that are an affine combination of a constraint's expectation and standard deviation \cite{Zhou2013, Xu2019, Farina2015, paulson2017stochastic}. These inequalities only require knowledge of the expectation and the standard deviation, which can be easily calculated for linear constraints. However, reliance on these inequalities typically provides quite conservative bounds \cite{paulson2017stochastic}. 

Our approach also invokes concentration inequalities, and hence provides almost surely guarantees, but employs an inequality that is less conservative than those in \cite{Zhou2013, Xu2019, Farina2015, paulson2017stochastic}. We use the one-sided Vysochanskij–Petunin inequality \cite{Mercadier2021}, a refinement of Cantelli's inequality that is tailored to unimodal distributions. Although it has less generality than Cantelli's inequality, the one-sided Vysochanskij–Petunin inequality typically results in far less conservatism in the overapproximation. Indeed, its probabilistic bound is reduced by a factor of 5/9, as compared to the bound from Cantelli's inequality. Hence, we propose application of the one-sided Vysochanskij–Petunin inequality to chance constraint evaluation that arises in multi-vehicle planning problems: that is, in a) reaching a terminal target set and b) avoiding collision with obstacles in the environment as well as with other vehicles. The main drawback in our approach is the need for unimodality of each constraint, over the entire trajectory. Unimodality is assured for convex constraints in LTI systems for certain classes of disturbance processes (such as Gaussian, Laplacian, or uniform on a convex interval), however for other disturbance processes, unimodality must be validated empirically. 

The main contribution of this paper is a {\em closed-form} reformulation of chance constraints, for polytopic target sets and collision avoidance constraints, that is amenable to difference of convex programming solutions. Our approach is relevant for LTI systems with arbitrary distributions with finite moments, and with chance constraints that are unimodal. 

The paper is organized as follows. Section \ref{sec:prelim} provides mathematical preliminaries and formulates the optimization problem. Section \ref{sec:methods} derives the difference of convex functions optimization problem reformulation of the chance constraints. Section \ref{sec:results} demonstrates our approach on two multi-satellite rendezvous problems, and Section \ref{sec:conclusion} provides concluding remarks.

\section{Preliminaries and Problem Formulation} \label{sec:prelim}

\subsection{Mathematical Preliminaries}

We denote the interval that enumerates all natural numbers from $a$ to $b$, inclusively, as $\Nt{a}{b}$. We denote vectors with an arrow accent, as $\vec{x} \in \R^n$. Random variables are indicated with a bold case $\boldsymbol{x}$. For a random variable $\boldsymbol{x}$, we denote the expectation as $\ex{\boldsymbol{x}}$, variance as $\var{\boldsymbol{x}}$, and standard deviation as $\std{\boldsymbol{x}}$. For a vector input, $\var{\cdot}$ will reference the variance-covariance matrix of the random vector. For two random variables, $\boldsymbol{x}$ and $\boldsymbol{y}$, $\cov{\boldsymbol{x}}{\boldsymbol{y}}$ denotes the covariance between the two variables. We denote the 2-norm of a matrix or vector by $\|\cdot\|$. For a matrix $A$, $\tr{A}$ will denote the trace of $A$. Last, we denote a block diagonal matrix with elements $A_1,A_2,\dots,A_q$ as $\mathrm{diag}(A_1,A_2,\dots,A_q)$.

\subsection{Problem Formulation}

\begin{figure}
    \centering
    \includegraphics[width=0.7\columnwidth]{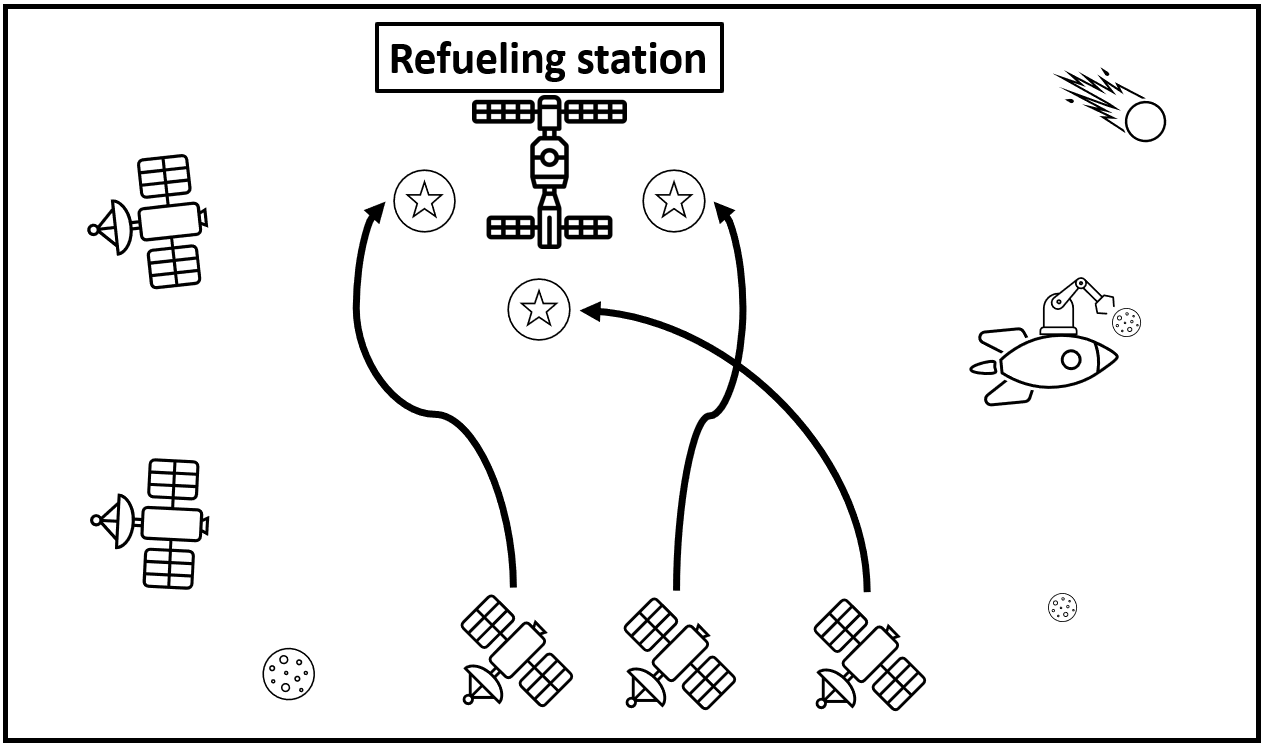}
    \caption{A scenario in which three satellites need to rendezvous with a refueling station while avoiding each other, other spacecraft, scientific instruments, and debris.}
    \label{fig:demo}
\end{figure}

Consider a scenario, such as the one shown in Figure \ref{fig:demo}, in which three satellites rendezvous with a refueling station while avoiding each other, other spacecraft, and debris. With potentially non-Gaussian disturbances corrupting the satellite dynamics, we seek to synthesize a controller to construct an optimal rendezvous maneuver that meets probabilistic target set and collision avoidance constraints.

We presume the evolution of $N_{v}$ vehicles are governed by the discrete-time LTI system,
\begin{equation}
    \bvec{x}_i(k+1) = A \bvec{x}_i(k) + B \vec{u}_i(k) + \bvec{w}_i(k) \label{eq:system}
\end{equation}
with state $\bvec{x}_i(k) \in \mathcal{X} \subseteq \R^n$, input $\vec{u}_i(k) \in \mathcal{U} \subseteq \R^m$,  $\bvec{w}_i(k) \in \R^n$ that follows an arbitrary but known disturbance, and initial condition $\vec{x}(0)$. We presume the initial conditions, $\vec{x}(0)$, are known, the bounded control authority, $\mathcal{U}$, is a convex polytope, and that the system evolves over a finite time horizon of $N \in \N$ steps. We presume each disturbance, $\bvec{w}_i(k)$, has probability space $(\Omega, \mathcal{B}(\Omega), \P_{\bvec{w}_i (k)})$ with outcomes $\Omega$, Borel $\sigma$-algebra $\mathcal{B}(\Omega)$, and probability measure $\P_{\bvec{w}_i (k)}$ \cite{casella2002}.

We write the dynamics at time $k$ as an affine sum of the initial condition and the concatenated control sequence and disturbance,
\begin{equation} \label{eq:lin_dynamics}
    \bvec{x}_i(k) = A^k \vec{x}_i(0) + \mathcal{C}(k) \vec{U}_i + \mathcal{D}(k) \bvec{W}_i
\end{equation}
with 
\begin{subequations}
\begin{alignat}{2}
    \vec{U}_i =& \left[ \vec{u}_i(0)^\top \ \ldots \ \vec{u}_i(N-1)^\top \right]^\top &&\in \mathcal{U}^{N} \\
    \bvec{W}_i =& \left[ \bvec{w}_i(0)^\top \ \ldots \ \bvec{w}_i(N-1)^\top \right]^\top &&\in \mathbb{R}^{Nn} \\
    \mathcal{C}(k) = & \left[ A^{k-1}B \ \ldots \ AB \ B \ 0_{n \times (N-k)m} \right] && \in \R^{n \times Nm} \\
    \mathcal{D}(k) = & \left[ A^{k-1} \ \ldots \ A \ I_n \ 0_{n \times (N-k)n} \right] && \in \R^{n \times Nn}
\end{alignat}
\end{subequations}    

We seek to minimize a convex performance objective $J: \mathcal{X}^{N \times N_v} \times \mathcal{U}^{N \times N_v} \rightarrow \R$. We presume desired polytopic target sets that each vehicle must reach, known and static obstacles that each vehicle must avoid, as well as the need for collision avoidance between each pair of vehicles, all with desired likelihoods,
\begin{subequations}\label{eq:constraints}
    \begin{align}
    \pr{\bigcap_{i=1}^{N_{v}}  \bigcap_{k=1}^{N} \bvec{x}_i(k)  \in  \mathcal{T}_{i}(k)} & \geq  1\!-\!\alpha \label{eq:constraint_t}\\
    \pr{\bigcap_{i=1}^{N_{v}}  \bigcap_{k=1}^{N} \| S (\bvec{x}_i(k) \!-\! \vec{o}(k)) \| \geq r} &\geq 1\!-\!\beta \label{eq:constraint_o}\\
    \pr{ \bigcap_{i=1}^{N_v-1} \bigcap_{j=i+1}^{N_v}  \bigcap_{k=1}^{N} \| S (\bvec{x}_i(k) \!-\!  \bvec{x}_j (k)) \| \geq r } &\geq 1\!-\!\gamma \label{eq:constraint_r}
    \end{align}
\end{subequations}
We presume convex, compact, and polytopic sets $ \mathcal{T}_{i}(k) \subseteq \R^n$, positive semi-definite and diagonal matrix $S \in \R^{q \times n}$, positive scalar $r \in \R_+$, non-random object locations $\vec{o}(k) \in \R^n$, and probabilistic violation thresholds $\alpha, \beta, \gamma \in (0,1/6)$. The probabilistic violation thresholds are restricted as a condition for optimally of the solutions. Here, $S$ is designed to extract the position of the vehicle from the state vector. 

\begin{mydefn}[Reverse convex constraint] \label{def:reverse-convex}
A reverse convex constraint is the complement of a convex constraint, that is, $f(x) \geq c$ for a convex function $f: \R^n \rightarrow \R$ and a scalar $c \in \R$.
\end{mydefn}

Note that the collision avoidance constraints inside the probability functions are reverse-convex as per Definition \ref{def:reverse-convex}. 

We seek to solve the following optimization problem.
\begin{subequations}\label{prob:big_prob_eq}
    \begin{align}
        \underset{\vec{U}_1, \dots, \vec{U}_{N_{v}}}{\mathrm{minimize}} \quad & J\left(
        \bvec{X}_1, \ldots, \bvec{X}_{N_{v}},  \vec{U}_1, \dots, \vec{U}_{N_{v}}\right)  \\
        \mathrm{subject\ to} \quad  & \vec{U}_1, \dots, \vec{U}_{N_{v}} \in  \mathcal U^N,  \\
        & \text{Dynamics } \eqref{eq:lin_dynamics} \text{ with }
        \vec{x}_1(0), \dots, \vec{x}_{N_{v}}(0)\\
        & \text{Probabilistic constraints  \eqref{eq:constraints}} \label{prob:initial_eq_prob_constraints} 
    \end{align}
\end{subequations}
where $\bvec{X}_i = \begin{bmatrix} \bvec{x}_i^{\top}(1) & \ldots & \bvec{x}_i^{\top}(N) \end{bmatrix}^{\top}$ is the concatenated state vector for vehicle $i$.

For this problem to be tractable for arbitrary disturbances, we make several key assumptions about the disturbance and its resulting impact on the constraints.

\begin{assm} \label{assm:vec_ind}
Disturbance vectors, $\bvec{W}_i$, are all pairwise independent. Hence, for any $i$ and $j$, where $i \neq j$, the joint CDF, $\Phi_{\bvec{W}_i, \bvec{W}_j}(\vec{a}, \vec{b})$ can be factored into the product of the marginal CDFs, $\Phi_{\bvec{W}_i}(\vec{a})$ and $\Phi_{\bvec{W}_j}(\vec{b})$. So, $\Phi_{\bvec{W}_i, \bvec{W}_j}(\vec{a}, \vec{b}) = \Phi_{\bvec{W}_i}(\vec{a}) \Phi_{\bvec{W}_j}(\vec{b})$.
\end{assm}
\begin{assm} \label{assm:elem_ind} 
All components of the disturbance vector, $\bvec{W}_i = \begin{bmatrix}\boldsymbol{w}_{i1} & \boldsymbol{w}_{i2} & \dots & \boldsymbol{w}_{iNn}\end{bmatrix}$, are mutually independent. Hence, for any set of unique integers $\mathbb{S} \subseteq \Nt{1}{Nn}$, the subset $\{ \boldsymbol{w}_{ij} | j \in \mathbb{S} \}$ has a joint CDF $\Phi_{\{ \boldsymbol{w}_{ij} | j \in \mathbb{S}\}}(\cdot, \ldots, \cdot)$ can be factored into the product of the marginal CDFs, $\Phi_{\boldsymbol{w}_{ij}}(\cdot)$ for $j \in \mathbb{S}$. So, $\Phi_{\{ \boldsymbol{w}_{ij} | j \in \mathbb{S}\}}(\cdot, \ldots, \cdot) = \prod_{j \in \mathbb{S}} \Phi_{\boldsymbol{w}_{ij}}(\cdot)$.
\end{assm}
\begin{assm} \label{assm:moments}
Each component of the disturbance vector, $\bvec{W}_i = \begin{bmatrix}\boldsymbol{w}_{i1} & \boldsymbol{w}_{i2} & \dots & \boldsymbol{w}_{iNn}\end{bmatrix}$, has finite and well defined moments at least up to the fourth order,  $\ex{\boldsymbol{w}_{ij}^p} < \infty$ for $p \in \Nt{1}{4}$.
\end{assm}

Statistically, pairwise and mutual independence can be assumed in many cases without much consequence as most multivariate distributions can be constructed in this manner. However, the multivariate Cauchy and the multivariate $t$ are the most prominent examples that cannot meet Assumption \ref{assm:elem_ind} as elements are not independent by construction. In many ways, Assumption \ref{assm:elem_ind} is the most restrictive of these assumptions as many physical phenomena may not disturb each state independently. Assumption \ref{assm:moments} is easily met as most distributions have analytic expressions for moments. 

Lastly, we consider the impact of $\bvec{W}_i$ on the chance constraints in \eqref{eq:constraints}. 

\begin{mydefn}[Unimodal Distribution \cite{Bertin1997}]\label{def:unimodal}
A unimodal distribution is a distribution whose CDF is convex in the region $(-\infty, a)$ and concave in the region $(a, \infty)$ for some $a \in \R$.
\end{mydefn}

\begin{mydefn}[Strong Unimodal Distribution \cite{Bertin1997}] \label{defn:strong_unimodal}
A strong unimodal distribution is one in which unimodality is preserved by convolution. That is, for two independent unimodal random variables, $\boldsymbol{y}$ and $\boldsymbol{z}$, the random variable $\boldsymbol{y} + \boldsymbol{z}$ is also unimodal.
\end{mydefn}

\begin{figure}
    \centering
    \includegraphics[width=0.9\columnwidth]{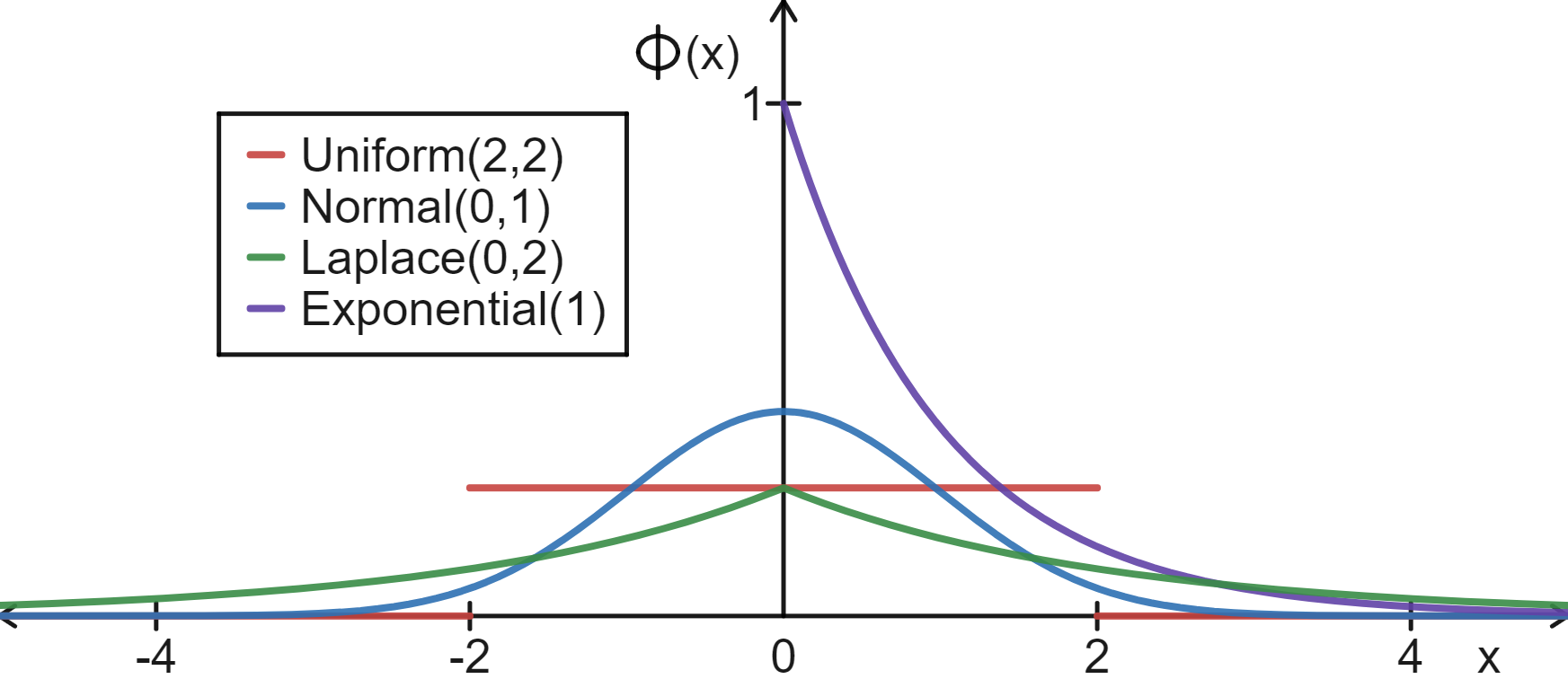}
    \includegraphics[width=0.9\columnwidth]{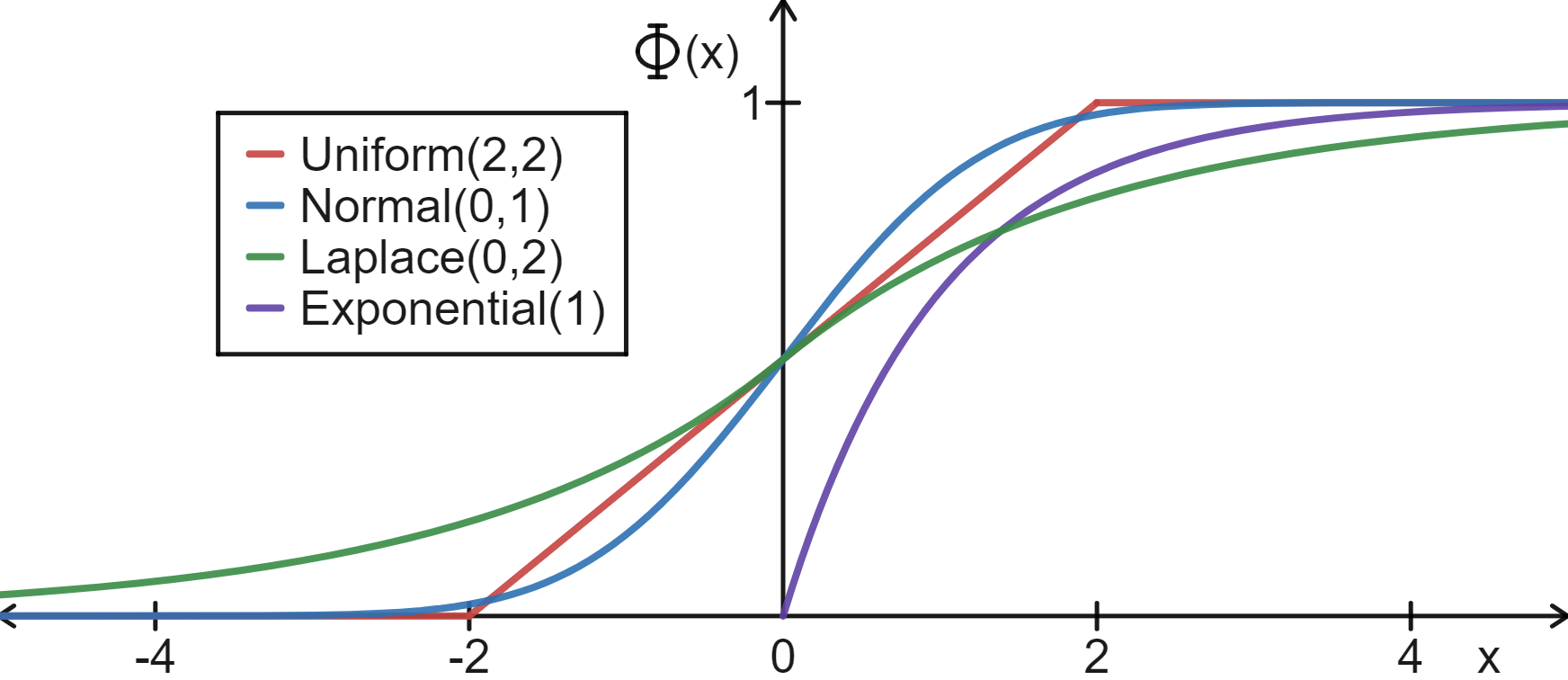}
    \caption{PDFs (top) and CDFs (bottom) of unimodal distributions as per Definition \ref{def:unimodal} with the mode at 0. Each of the distributions shown here have a log concave PDF, which in turn implies a log concave CDF. Log concavity of the PDF assures strong unimodality as per Definition \ref{defn:strong_unimodal}. }
    \label{fig:unimodal}
\end{figure}

\begin{assm}\label{assm:unimodal}
The distribution that describes each probabilistic constraint in \eqref{eq:constraints} is marginally unimodal.
\end{assm}

Assumption \ref{assm:unimodal} is required such that we can develop bounds on the chance constraint probabilities. In rare cases, unimodality can be verified analytically by properties of strong unimodality. For example, Gaussian or exponential random variables are strong unimodal and any affine summation of these random variables will always be unimodal. One method to check for strong unimodality is to establish that the probability density function (PDF) is log concave as all distributions that are strong unimodal also have a log concave PDF per the Theorem of Ibragimov \cite{Ibragimov1956}. Figure \ref{fig:unimodal} graphs the PDF and CDF of several common strong unimodal distributions with PDFs that are easy to show are log concave. As unimodality can be challenging to show analytically, the easiest method to validate unimodality is empirically. By numerically evaluating the empirical cumulative distribution function with a large enough sample size (we recommend at least on the order of $10^4$ samples), one can validate unimodality in terms of Definition \ref{def:unimodal} via Algorithm \ref{algo:unimodal} in Appendix \ref{appx:unimodal}. 

\begin{prob} \label{prob:initial}
    Under Assumptions \ref{assm:vec_ind}-\ref{assm:unimodal}, solve the stochastic optimization problem \eqref{prob:big_prob_eq} with probabilistic violation thresholds $\alpha$, $\beta$, and $\gamma$ for open loop controllers $\vec{U}_1, \dots, \vec{U}_{N_{v}} \in  \mathcal U^N$.
\end{prob} 

The main challenge in solving Problem \ref{prob:initial} is assuring \eqref{prob:initial_eq_prob_constraints}. In this form, assuring \eqref{prob:initial_eq_prob_constraints} requires the evaluation of high dimensional and frequently intractable integrals. Additionally, even if these integrals could be evaluated and closed forms could be found, the collision avoidance constraints \eqref{eq:constraint_o}-\eqref{eq:constraint_r} would still be reverse convex. 

\section{Methods} \label{sec:methods}

Our approach to solve Problem \ref{prob:initial} involves reformulating each chance constraint as an affine summation of the constraint's expectation and standard deviation, i.e., $\ex{\| S (\bvec{x}_i(k) \!-\!  \bvec{x}_j (k)) \|}$ and $\std{\| S (\bvec{x}_i(k) \!-\!  \bvec{x}_j (k)) \|}$, respectively for the collision avoidance constraint. This form is amenable to the one-sided Vysochanskij–Petunin inequality \cite{Mercadier2021}, which allows for almost surely guarantees of chance constraint satisfaction.

\begin{thm}[One-sided Vysochanskij–Petunin Inequality \cite{Mercadier2021}] \label{thm:osvpi}
Let $\boldsymbol{x}$ be a real valued unimodal random variable with finite expectation $\ex{\boldsymbol{x}}$ and finite, non-zero standard deviation $\std{\boldsymbol{x}}$. Then, for $\lambda > \sqrt{5/3}$, 
\end{thm}
\begin{equation} \label{eq:vp_1}
    \pr{\boldsymbol{x} - \ex{\boldsymbol{x}}  \geq  \lambda \std{\boldsymbol{x}}} \leq \frac{4}{9(\lambda^2+1)}
\end{equation}

By applying \eqref{eq:vp_1} to the random variable $-\boldsymbol{x}$, we get the lower tail bound
\begin{equation}
    \pr{\boldsymbol{x} -  \ex{\boldsymbol{x}} \leq - \lambda \std{\boldsymbol{x}}} \leq \frac{4}{9(\lambda^2+1)} \label{eq:vp_2}
\end{equation}

The one-sided Vysochanskij–Petunin inequality is applicable only to unimodal distributions. It is based on Gauss's inequality, which provides a bound for one sided tail probabilities of a unimodal random variable to be sufficiently far away from the expectation. Specifically, the bound encompasses values at least $\lambda$ standard deviations away from the mean. 

We first make use of \eqref{eq:vp_1} and \eqref{eq:vp_2} to bound the chance constraint probabilities based on an affine summation of the expectation and standard deviation.

\subsection{Polytopic Target Set Constraint} \label{ssec:target_reform}

First, consider the reformulation of \eqref{eq:constraint_t}. Without loss of generality, we presume $N_v=1$ and $N=1$ for brevity. The polytope $\mathcal{T}_{i}(k)$ can be written as the intersection of $N_{Tik}$ half-space inequalities,
\begin{equation}
    \pr{\bvec{x}_i(k) \in \mathcal{T}_{i}(k)} =  \pr{ \bigcap_{j=1}^{N_{Tik}} \vec{G}_{ijk} \bvec{x}_i(k) \leq h_{ijk}}
\end{equation}
where $\vec{G}_{ijk} \in \R^n$ and $h_{ijk} \in \R$. We take the complement and employ Boole's inequality to separate the combined chance constraints into a series of individual chance constraints,
\begin{subequations}
\begin{align}
    \pr{\boldsymbol{x}_i(k) \not\in \mathcal{T}_{i}(k) } = & \; \pr{ \bigcup_{j=1}^{N_{Tik}} \vec{G}_{ijk} \bvec{x}_i(k) \geq h_{ijk}} \\
    \leq & \; \sum_{j=1}^{N_{Tik}} \pr{\vec{G}_{ijk} \bvec{x}_i(k) \geq h_{ijk}}
\end{align}
\end{subequations}
Using the approach in \cite{ono2008iterative}, we introduce variables $\omega_{ijk}$ to allocate risk to each of the individual chance constraints,
\begin{subequations}\label{eq:quantile_reform_new_var}
\begin{align}
     \pr{\vec{G}_{ijk} \bvec{x}_i(k) \geq h_{ijk}} &\leq \omega_{ijk} \label{eq:quantile_orig} \\
     \sum_{j=1}^{N_{Tik}} \omega_{ijk} &\leq \alpha \label{eq:quantile_reform_new_var_2}\\
     \omega_{ijk} & \geq 0 \label{eq:quantile_reform_new_var_3}
\end{align}
\end{subequations}
To find a solution to \eqref{eq:quantile_reform_new_var}, we need to find an appropriate value for $\omega_{ijk}$. To that end, we add an additional constraint 
\begin{equation} \label{eq:add_target}
    \ex{\vec{G}_{ijk}\bvec{x}_i(k)} + \lambda_{ijk} \std{\vec{G}_{ijk} \bvec{x}_i(k)} \leq  h_{ijk}
\end{equation}
to \eqref{eq:quantile_reform_new_var}. Enforcement of \eqref{eq:add_target} allows us to write \eqref{eq:quantile_orig} as 
\begin{equation} \label{eq:enforce_target}
\begin{split}
     & \pr{\vec{G}_{ijk} \bvec{x}_i(k) \geq h_{ijk}}  \\
     & \ \leq \pr{\vec{G}_{ijk} \bvec{x}_i(k) \geq \ex{\vec{G}_{ijk} \bvec{x}_i(k)} + \lambda_{ijk} \std{\vec{G}_{ijk} \bvec{x}_i(k) }} \\
     & \ \leq \omega_{ijk}
\end{split}
\end{equation} 
Then, by Assumption \ref{assm:unimodal} and Theorem \ref{thm:osvpi}, we can substitute $\omega_{ijk}$ with $\frac{4}{9(\lambda_{ijk}^2+1)}$ and change the risk allocation variable from $\omega_{ijk}$ to $\lambda_{ijk}$. Further, enforcement of \eqref{eq:add_target} makes \eqref{eq:quantile_orig}, and by extension \eqref{eq:enforce_target}, an unnecessary intermediary step between \eqref{eq:add_target} and \eqref{eq:quantile_reform_new_var_3}. Hence, we can remove \eqref{eq:quantile_orig} from the system of equations to solve and write \eqref{eq:quantile_reform_new_var}-\eqref{eq:enforce_target} as 
\begin{subequations}\label{eq:target_constraint}
\begin{align}
    \ex{\vec{G}_{ijk}\bvec{x}_i(k)} \!+\! \lambda_{ijk} \std{\vec{G}_{ijk} \bvec{x}_i(k)} \leq & \; h_{ijk}  \label{eq:target_reform}\\ 
    \sum_{j=1}^{N_{Tik}} \frac{4}{9(\lambda_{ijk}^2+1)} \leq & \; \alpha & \label{eq:target_lambda}\\
    \lambda_{ijk} \geq & \; \sqrt{\frac{5}{3}}  \label{eq:lambda_restrict}
\end{align}
\end{subequations}
which is enumerated over the indices, $i$, $j$, and $k$.

\begin{lem} \label{lem:target_satisfy}
For the controllers $\vec{U}_1, \dots, \vec{U}_{N_{v}}$, if there exists risk allocation variables $\lambda_{ijk}$ satisfying \eqref{eq:target_constraint} for constraints \eqref{eq:constraint_t}, then $\vec{U}_1, \dots, \vec{U}_{N_{v}}$ satisfy \eqref{prob:initial_eq_prob_constraints}.
\end{lem}

\begin{IEEEproof}
Satisfaction of \eqref{eq:target_reform} implies \eqref{eq:enforce_target} holds. The Vysochanskij–Petunin inequality upper bounds \eqref{eq:enforce_target}. Boole's inequality and De Morgan's law guarantee that if \eqref{eq:target_lambda} holds then \eqref{prob:initial_eq_prob_constraints} is satisfied.
\end{IEEEproof}

Lastly, we show that the constraint reformulation \eqref{eq:target_constraint} will always be convex.

\begin{lem} \label{lem:target_convex}
The constraint \eqref{eq:target_constraint} is convex in $\vec{U}_i$ and in $(\lambda_{i1k}, \dots, \lambda_{ipk})$
\end{lem}
\begin{IEEEproof}
We start by exploiting the properties of the expectation and variance operator to write \eqref{eq:target_reform} as
\begin{equation}
\begin{split}
    & \vec{G}_{ijk} \left(A^k \vec{x}_i(0) + \mathcal{C}(k) \vec{U}_i + \mathcal{D}(k)\ex{\bvec{W}_i} \right) \\
    & \quad + \lambda_{ijk} \sqrt{ \vec{G}_{ijk}^\top \mathcal{D}^{\top}(k) \var{\bvec{W}_i} \mathcal{D}(k) \vec{G}_{ijk} } \leq h_{ijk}    
\end{split}
\end{equation}
which is affine, and hence convex, in $\vec{U}_i$ and $\lambda_{ijk}$. Then
\begin{equation}
    \frac{\partial^2}{\partial \lambda_{ijk}^2}\frac{4}{9(\lambda_{ijk}^2+1)} = -\frac{8\left(-3\lambda_{ijk}^2+1\right)}{9\left(\lambda_{ijk}^2+1\right)^3}
\end{equation}
which is positive, and hence convex, when $\lambda_{ijk} \geq 3^{-1/2}$. Hence, with the restriction \eqref{eq:lambda_restrict}, \eqref{eq:target_lambda} is a convex constraint. Thus, the set over which $\lambda_{i1k}, \dots, \lambda_{iqk}$ is optimized is convex. Further, in the problem formulation we defined the control authority to be a closed and convex set. Hence, we can conclude the chance constraint reformulation \eqref{eq:target_constraint} is convex.
\end{IEEEproof}

\subsection{2-Norm Based Collision Avoidance Constraints} \label{ssec:collision_reform}

Next, consider the reformulation of the constraints \eqref{eq:constraint_o}-\eqref{eq:constraint_r}. Here, we will derive the reformulation for \eqref{eq:constraint_r}, but the reformulation of \eqref{eq:constraint_o} is nearly identical. Without loss of generality, let 
\begin{subequations}
\begin{align}
    \vec{z} = & \; S A^k (\vec{x}_i(0)-\vec{x}_j(0)) + S \mathcal{C}(k) (\vec{U}_i-\vec{U}_j) \\
    \bvec{z} = & \; S  \mathcal{D}(k)( \bvec{W}_i- \bvec{W}_j)
\end{align}
\end{subequations}
be the non-stochastic and stochastic element of $S(\bvec{x}_i(k) - \bvec{x}_j (k))$ from \eqref{eq:constraint_r}, respectively. Then, we can write the norm as
\begin{equation}
    \|S(\bvec{x}_i(k) - \bvec{x}_j (k))\| = \| \vec{z} + \bvec{z} \|
\end{equation}
We start by observing 
\begin{equation} \label{eq:norm_square}
\begin{split}
    &\pr{ \bigcap_{i=1}^{N_v-1} \bigcap_{j=i+1}^{N_v}  \bigcap_{k=1}^{N} \| \vec{z} + \bvec{z} \| \geq r } \\
    & \ = \pr{\bigcap_{i=1}^{N_v-1} \bigcap_{j=i+1}^{N_v}  \bigcap_{k=1}^{N} \| \vec{z} + \bvec{z} \|^2 \geq r^2 }    
\end{split}
\end{equation}
as the norm is non-negative. Thus, we can write the 2-norm constraint as
\begin{equation} \label{eq:square_norm}
    \pr{\bigcap_{i=1}^{N_v-1} \bigcap_{j=i+1}^{N_v}  \bigcap_{k=1}^{N} \| \vec{z} + \bvec{z} \|^2 \geq r^2} \geq 1-\gamma
\end{equation}

By taking the complement and applying Boole's inequality,
\begin{equation}
\begin{split}
    &\pr{\bigcup_{i=1}^{N_v-1} \bigcup_{j=i+1}^{N_v}  \bigcup_{k=1}^{N} \| \vec{z} + \bvec{z} \|^2 \leq r^2 } \\
    & \ \leq \sum_{i=1}^{N_v-1} \sum_{j=i+1}^{N_v}  \sum_{k=1}^{N} \pr{ \| \vec{z} + \bvec{z} \|^2 \leq r^2 }    
\end{split}
\end{equation}
Using the approach in \cite{ono2008iterative}, we introduce risk variables $\omega_{ijk}$ to allocate risk to each of the individual probabilities
\begin{subequations}\label{eq:reform_new_var}
\begin{align}
   \pr{ \| \vec{z} + \bvec{z} \|^2 \leq r^2 } & \leq \omega_{ijk}  \\
  \sum_{i=1}^{N_v-1} \sum_{j=i+1}^{N_v}  \sum_{k=1}^{N} \omega_{ijk} &\leq \gamma \\
  \omega_{ijk} & \geq 0 
\end{align}
\end{subequations}

In a similar fashion to Section \ref{ssec:target_reform}, we add an additional constraint based on the expectation and standard deviation of $\| \vec{z} + \bvec{z} \|^2$ to \eqref{eq:reform_new_var} such that the constraint becomes
\begin{subequations} \label{eq:norm_reform_p1}
\begin{align}
    \pr{\| \vec{z} + \bvec{z} \|^2 \leq r^2} \leq & \; \omega_{ijk} \label{eq:norm_reform_p1_1} \\
    \ex{\| \vec{z} + \bvec{z} \|^2} - \lambda_{ijk} \std{\| \vec{z} + \bvec{z} \|^2}  \geq & \; r^2 \label{eq:norm_reform_p1_2} \\
    \sum_{i=1}^{N_v-1} \sum_{j=i+1}^{N_v}  \sum_{k=1}^{N} \omega_{ijk} \leq & \; \gamma \\
    \omega_{ijk} \geq &\;  0 
\end{align}
\end{subequations}
By enforcing \eqref{eq:norm_reform_p1_2}, we can write \eqref{eq:norm_reform_p1_1} as 
\begin{equation} \label{eq:norm_bound}
\begin{split}
    & \pr{\| \vec{z} + \bvec{z} \|^2 \leq r^2}   \\
    & \ \leq \pr{
        \begin{subarray}{l}
            \| \vec{z} + \bvec{z} \|^2 \\
            \ \leq \; \ex{\| \vec{z} + \bvec{z} \|^2} - \lambda_{ijk}\std{\| \vec{z} + \bvec{z} \|^2} 
        \end{subarray}} \\
    & \ \leq \omega_{ijk}
\end{split}
\end{equation}
From Assumption \ref{assm:unimodal} and Theorem \ref{thm:osvpi}, we know that \eqref{eq:norm_bound} is upper bounded as per \eqref{eq:vp_2}. Hence, we can use the substitution 
\begin{equation}
    \omega_{ijk} = \frac{4}{9(\lambda_{ijk}^2+1)} 
\end{equation}
and determine the value for $\lambda_{ijk}$ in terms of $\omega_{ijk}$,
\begin{equation}
    \lambda_{ijk} = \sqrt{\frac{4}{9\omega_{ijk}}-1}
\end{equation}
so long as $\lambda \geq \sqrt{5/3}$. This implies $\omega_{ijk} \leq 1/6$ is a necessary restriction on $\omega_{ijk}$. As $\gamma < 1/6$, any solution will require that $\omega_{ijk} < 1/6$. Then, we can write \eqref{eq:norm_reform_p1} as
\begin{subequations} \label{eq:norm_reform_p2} 
\begin{align}
    \pr{\| \vec{z} + \bvec{z} \|^2 \leq r^2} \leq & \; \omega_{ijk} \label{eq:norm_reform_p2_1}\\
    \ex{\| \vec{z} \!+\! \bvec{z} \|^2} \!-\! \sqrt{\frac{4}{9\omega_{ijk}}\!-\!1} \cdot  \std{ \| \vec{z} \!+\! \bvec{z} \|^2}  \geq & \; r^2  \label{eq:norm_reform_p2_2}\\
    \sum_{i=1}^{N_v-1} \sum_{j=i+1}^{N_v}  \sum_{k=1}^{N} \omega_{ijk} \leq &\; \gamma \\
    \omega_{ijk} \in &\; (0,1/6)
\end{align}
\end{subequations}
Since Theorem \ref{thm:osvpi} guarantees that satisfaction of \eqref{eq:norm_reform_p2_2} also satisfies \eqref{eq:norm_reform_p2_1} for any value $\gamma \in (0,1/6)$, \eqref{eq:norm_reform_p2_1} is redundant and can be removed. The constraint is then
\begin{subequations} \label{eq:norm_reform_p2.5}
\begin{align}
    \ex{\| \vec{z} \!+\! \bvec{z} \|^2} \!-\! \sqrt{\frac{4}{9\omega_{ijk}}\!-\!1} \cdot  \std{ \| \vec{z} \!+\! \bvec{z} \|^2}  \geq & \; r^2  \label{eq:norm_reform_p2.5_2}\\
    \sum_{i=1}^{N_v-1} \sum_{j=i+1}^{N_v}  \sum_{k=1}^{N} \omega_{ijk} \leq &\; \gamma \\
    \omega_{ijk} \in &\; (0,1/6)
\end{align}
\end{subequations}
Note that \eqref{eq:norm_reform_p2.5_2} is a biconvex constraint \cite{Gorski2007}. For known risk allocation values $\omega_{ijk}$, the final constraint is,
\begin{equation} \label{eq:norm_reform_p3}
        \ex{\| \vec{z} + \bvec{z} \|^2} - \sqrt{\frac{4}{9\tilde{\omega}_{ijk}}-1} \cdot \std{ \| \vec{z} + \bvec{z} \|^2}  \geq r^2
\end{equation}

\begin{lem} \label{lem:collision_satisfy}
If the controller $\vec{U}_1, \dots, \vec{U}_v$, satisfies \eqref{eq:norm_reform_p3} for constraints \eqref{eq:constraint_o}-\eqref{eq:constraint_r}, then  $\vec{U}_1, \dots, \vec{U}_v$ satisfy \eqref{eq:constraints}.
\end{lem}

\begin{IEEEproof}
Satisfaction of \eqref{eq:norm_reform_p3} implies \eqref{eq:norm_bound} is satisfied for $\lambda_{ijk} = \sqrt{\frac{4}{9\tilde{\omega}_{ijk}}-1}$. Theorem \ref{thm:osvpi} guarantees satisfaction of \eqref{eq:constraints}. 
\end{IEEEproof}

Next, we find the expanded form of \eqref{eq:norm_reform_p3} and show that the constraint is always a difference of convex function constraint. 

\begin{mydefn}[Difference of Convex Functions Constraint] \label{defn:dc}
A difference of convex functions constraint has the form
\begin{equation}\label{eq:dc_def}
   f(\vec{x})-g(\vec{x}) \leq 0 
\end{equation}
in which $f, g: \R^n \rightarrow \R$ are convex functions for $\vec{x} \in \R^n$.
\end{mydefn}

\begin{lem}
The constraint \eqref{eq:norm_reform_p3} is a difference of convex function constraint in $\vec{U}_i$ for the constraint \eqref{eq:constraint_o} in $\vec{U}_i-\vec{U}_j$ for the constraint \eqref{eq:constraint_r}. 
\end{lem}
\begin{IEEEproof}
We first find the expectation and variance of the norm. To find the expectation, we expand the norm,
\begin{subequations} \label{eq:norm_expect}
\begin{align}
    \ex{\|\vec{z} + \bvec{z} \|^2}
    & \ = \ex{\vec{z}^{\top}\vec{z} + 2 \vec{z}^{\top} \bvec{z} + \bvec{z}^{\top}\bvec{z}} \\
    & \ =  \vec{z}^{\top}\vec{z} + 2 \vec{z}^{\top} \ex{\bvec{z}} + \ex{\bvec{z}^{\top}\bvec{z}} \\
    & \ = \left\|  
        \begin{bmatrix} 
            I_q &  \ex{\bvec{z}}  \\
             \ex{\bvec{z}}^{\top} & \ex{\bvec{z}^{\top}\bvec{z}}
        \end{bmatrix}^{\frac{1}{2}}
        \begin{bmatrix} 
            \vec{z}\\
            1
        \end{bmatrix}
    \right\|^2 \label{eq:norm_exp_norm}
\end{align}
\end{subequations}
Remember $q$ is the dimension of the matrix $S$ designed to extract the position elements of the state. Here, $\ex{\|\vec{z} + \bvec{z} \|^2}$ is the squared norm of a vector matrix product. Hence, the expectation is convex. we compute the variance in a similar manner to the expectation,
\begin{subequations} \label{eq:var_expand_collision}
\begin{align}
    & \var{\| \vec{z} + \bvec{z} \|^2} \\
    & \ = \var{ \vec{z}^{\top}\vec{z} + 2 \vec{z}^{\top} \bvec{z} + \bvec{z}^{\top}\bvec{z}} \\
    & \ = \var{2 \vec{z}^{\top} \bvec{z}} + 2\cov{2\vec{z}^{\top} \bvec{z}}{ \bvec{z}^{\top}\bvec{z}} + \var{\bvec{z}^{\top}\bvec{z}}  \\
    & \ = 4 \vec{z}^{\top} \var{\bvec{z}} \vec{z}\! +\! 4 \vec{z}^{\top} \cov{\bvec{z}}{\bvec{z}^{\top}\bvec{z}} \!+\! \var{\bvec{z}^{\top}\bvec{z}}
\end{align}
\end{subequations}
where
\begin{equation}
    \cov{\bvec{z}}{\bvec{z}^{\top}\bvec{z}} = \ex{\bvec{z} \bvec{z}^{\top}\bvec{z}} - \ex{\bvec{z}} \ex{\bvec{z}^{\top}\bvec{z}}
\end{equation}
Assumptions \ref{assm:vec_ind}-\ref{assm:moments} guarantees a closed form for \eqref{eq:var_expand_collision}. Thus, we can write the standard deviation as the 2-norm
\begin{equation} \label{eq:var_IEEEproof}
\begin{split}
    & \std{\| \vec{z} + \bvec{z} \|^2} \\
    & \ = \left\|  
        \begin{bmatrix} 
            4  \var{\bvec{z}} & 2\cov{\bvec{z}}{\bvec{z}^{\top}\bvec{z}}  \\
            2 \cov{\bvec{z}}{\bvec{z}^{\top}\bvec{z}}^{\top} & \var{\bvec{z}^{\top} \bvec{z}}
        \end{bmatrix}^{\frac{1}{2}}
        \begin{bmatrix} 
            \vec{z}\\
            1
        \end{bmatrix}
    \right\| 
\end{split}
\end{equation}
Since the standard deviation is the 2-norm of an affine function, the standard deviation is convex \cite{boyd_convex}. 
\begin{figure*}
\begin{equation}
    \left\|  
        \begin{bmatrix} 
            I_q &  \ex{\bvec{z}}  \\
             \ex{\bvec{z}}^{\top} & \ex{\bvec{z}^{\top}\bvec{z}}
        \end{bmatrix}^{\frac{1}{2}}
        \begin{bmatrix} 
            \vec{z}\\
            1
        \end{bmatrix}
    \right\|^2 - \sqrt{\frac{4}{9\tilde{\omega}_{ijk}}-1} \left\|
        \begin{bmatrix} 
            4  \var{\bvec{z}} & 2\cov{\bvec{z}}{\bvec{z}^{\top}\bvec{z}}  \\
            2 \cov{\bvec{z}}{\bvec{z}^{\top}\bvec{z}}^{\top} & \var{\bvec{z}^{\top} \bvec{z}}
        \end{bmatrix}^{\frac{1}{2}}
        \begin{bmatrix} 
            \vec{z}\\
            1
        \end{bmatrix}
    \right\|   \geq r^2 \label{eq:norm_reform_p4} 
\end{equation}
\hrulefill
\end{figure*}
We can now write \eqref{eq:norm_reform_p3} as \eqref{eq:norm_reform_p4}. By multiplying both sides of \eqref{eq:norm_reform_p4} by $-1$, \eqref{eq:norm_reform_p4} is the difference of two convex functions in the control as per Definition \ref{defn:dc}. 
\end{IEEEproof}

\subsection{Difference of Convex Functions Framework} \label{ssec:dc}

Combining the results from Sections \ref{ssec:target_reform} and \ref{ssec:collision_reform}, we obtain a new optimization problem.
\begin{subequations}\label{prob:big_prob_eq_2}
    \begin{align}
        \underset{\substack{\vec{U}_1, \dots, \vec{U}_{N_{v}} \\ \lambda_{ijk}}}{\mathrm{minimize}} \quad & J\left(
        \bvec{X}_1, \ldots, \bvec{X}_{N_{v}},  \vec{U}_1, \dots, \vec{U}_{N_{v}}\right) \label{eq:dc_cost} \\
        \mathrm{subject\ to} \quad  & \vec{U}_1, \dots, \vec{U}_{N_{v}} \in  \mathcal U^N,  \\
        & \text{Moments defined by dynamics \eqref{eq:lin_dynamics}} \label{eq:dc_moments} \\
        & \text{with initial conditions }
        \vec{x}_1(0), \dots, \vec{x}_{N_{v}}(0) \nonumber\\
        & \text{Constraints  \eqref{eq:target_constraint} and \eqref{eq:norm_reform_p4}} \label{prob:second_eq_prob_constraints} 
    \end{align}
\end{subequations}

\begin{reform} \label{prob:second}
    Under Assumptions \ref{assm:vec_ind}-\ref{assm:unimodal}, solve the stochastic optimization problem \eqref{prob:big_prob_eq_2} with probabilistic violation thresholds $\alpha$, $\beta$, and $\gamma$ for open loop controllers $\vec{U}_1, \dots, \vec{U}_{N_{v}} \in  \mathcal U^N$ and optimization parameters $\lambda_{ijk}$. 
\end{reform} 

\begin{lem}
Solutions to Reformulation \ref{prob:second} are conservative solutions to Problem \ref{prob:initial}.
\end{lem}
\begin{IEEEproof}
Lemmas \ref{lem:target_satisfy} and \ref{lem:collision_satisfy} guarantee the probabilistic constraints \eqref{eq:constraints} are satisfied. The equations \eqref{eq:vp_1}-\eqref{eq:vp_2} are always conservative. Hence, the reformulated constraints will be conservative with respect to the chance constraint. The expectation and variance terms in Reformulation \ref{prob:second} encompass and replace the dynamics used in Problem \ref{prob:initial}. The cost function and input constraints remain unchanged. 
\end{IEEEproof}

We note that \eqref{prob:big_prob_eq_2} is a difference of convex functions optimization problem. A difference of convex functions optimization problem has the form
\begin{equation} \label{eq:dc}
\begin{split}
    \underset{x}{\mathrm{minimize}} \quad & f_0(x)-g_0(x)  \\
    \mathrm{subject\ to} \quad  & f_i(x)-g_i(x) \leq 0 \quad \text{for } i \in \mathbb{N}  \\
\end{split}   
\end{equation}
in which $f_0, f_i(\cdot): \R^n \rightarrow \R$ and $g_0, g_i(\cdot): \R^n \rightarrow \R$ for $x \in \R^n$ are convex. While \eqref{eq:dc_cost}-\eqref{eq:dc_moments} are convex,  \eqref{prob:second_eq_prob_constraints} is difference of convex due to the constraint \eqref{eq:norm_reform_p4}. 

We employ the convex-concave procedure \cite{boyd_dc_2016} to solve \eqref{prob:big_prob_eq_2}. By taking a first order approximation of the expectation of the 2-norm in \eqref{eq:norm_reform_p4}, we can solve the difference of convex function optimization problem iteratively as a convex optimization problem. By updating the first order approximation at each iteration, the convex-concave procedure solves to a local optimum. Here, the first order approximation transforms the difference of convex function constraint \eqref{eq:norm_reform_p4} into the convex constraint \eqref{eq:norm_reform_p5} where the superscript $p$ indicated the value from the previous iteration’s solution. The main benefit of solving this problem with the convex-concave procedure is the first order approximation makes the constraint convex while maintaining the probabilistic assurances.
\begin{figure*}
    \begin{equation}
\begin{split}
    & \sqrt{\frac{4}{9\Tilde{\omega}_{ijk}}-1} \left\|  
        \begin{bmatrix} 
            4  \var{\bvec{z}} & 2\cov{\bvec{z}}{\bvec{z}^{\top}\bvec{z}}  \\
            2 \cov{\bvec{z}}{\bvec{z}^{\top}\bvec{z}}^{\top} & \var{\bvec{z}^{\top} \bvec{z}}
        \end{bmatrix}^{\frac{1}{2}}
        \begin{bmatrix} 
            \vec{z}\\
            1
        \end{bmatrix}
    \right\|  \\
    & \ -
   \underbrace{\left(\left\|
        \begin{bmatrix} 
            I_q &  \ex{\bvec{z}}  \\
             \ex{\bvec{z}}^{\top} & \ex{\bvec{z}^{\top}\bvec{z}}
        \end{bmatrix}^{\frac{1}{2}}
        \begin{bmatrix} 
            \vec{z}^p\\
            1
        \end{bmatrix}\right\|^2 +  2 \left(\vec{z}^p + \ex{\bvec{z}} \right)^{\top} S \mathcal{C}(k) \left(
            (\vec{U}_i -\vec{U}_j) - (\vec{U}_i^p
            + \vec{U}_j^p) \right)
        \right)}_{\text{First order approximation of }\ex{\|\vec{z} + \bvec{z} \|^2}\text{ based on previous iteration's solution.}}
     \leq -r^2 
\end{split}\label{eq:norm_reform_p5} 
\end{equation}
\hrulefill
\end{figure*}
Since feasibility of \eqref{eq:dc} is dependent on the feasibility of the initial conditions, we use slack variables to accommodate potentially infeasible initial conditions that can occur during the iterative process \cite{boyd_dc_2016,horst2000}. As we use a difference of convex functions optimization framework, Lemma \ref{lem:collision_satisfy} guarantees that any solution that is synthesized during iterative process will be a feasible but locally optimal solution.

\section{Results} \label{sec:results}

We demonstrate our method on a multi-satellite rendezvous problem with two different disturbances that impact the relative satellite dynamics. All computations were done on a 1.80GHz i7 processor with 16GB of RAM, using MATLAB, CVX \cite{cvx} and Gurobi \cite{gurobi}. Polytopic construction and plotting was done with MPT3 \cite{MPT3}. All code is available at \url{https://github.com/unm-hscl/shawnpriore-moment-control}.

Consider a scenario in which $N_v$ satellites, called the deputies, are stationed in geostationary Earth orbit, and tasked to rendezvous with a refueling spacecraft, called the chief. The satellites are tasked with reaching a new configuration represented by polytopic target sets. Each deputy must avoid other deputies while navigating to their respective target sets. The relative planar dynamics of each deputy, with respect to the position of the chief are described by the CWH equations \cite{wiesel1989_spaceflight}
\begin{subequations}
\begin{align}
\ddot x - 3 \omega^2 x - 2 \omega \dot y &= \frac{F_x}{m_c} \label{eq:cwh:a}\\
\ddot y + 2 \omega \dot x & = \frac{F_y}{m_c} \label{eq:cwh:b}
\end{align}   
\label{eq:cwh}
\end{subequations}
with input $\vec{u}_i = [ \begin{array}{ccc} F_x & F_y \end{array}]^\top$, and orbital rate $\omega = \sqrt{\frac{\mu}{R^3_0}}$, Earth's gravitational parameter $\mu$, orbital radius $R_0$, and mass of the deputy. We discretize \eqref{eq:cwh} under impulsive thrust assumptions, with sampling time $\Delta$t$=$ 60s, and insert a disturbance process that captures uncertainties in the model specification, so that dynamics of each deputy are described by  
\begin{equation}
    \bvec{x}_i(k+1) = A \bvec{x}_i(k) + B \vec{u}_i(k) + \bvec{w}_i(k)
\end{equation}
We assume that the disturbances adhere to Assumptions \ref{assm:vec_ind}-\ref{assm:elem_ind}.

\subsection{Exponential Disturbance} \label{ssec:exp_ex}

Exponential disturbances are the type of distribution that is a big motivator for our approach. They exist in real systems but very few methods can handle them. We presume, for the purpose of demonstration, that we have an exponential disturbance. This could occur because of inaccuracies in the impulsive thrust model, drag forces in low Earth orbit, or third body gravity.

The exponential distribution is defined as follows.

\begin{mydefn}[Exponential Distribution]
An exponential distribution is one which elicits the PDF
\begin{equation}
    \phi(x) = \lambda e^{-\lambda x}
\end{equation}
with rate parameter $\lambda>0$ and $x \geq 0$.
\end{mydefn}

The exponential distribution presents several challenges for existing methods. We define the following two distributions to analyze these challenges. 

\begin{mydefn}[Hypoexponential Distribution]
An hypoexponential distribution is one which elicits the PDF
\begin{equation}
    \phi(x) = -\vec{a} e^{x \Theta} \Theta \vec{1}
\end{equation}
with probability row vector $\vec{a}$, subgenerator matrix $\Theta$, and $x \geq 0$.
\end{mydefn}

\begin{mydefn}[Weibull Distribution]
An Weibull distribution is one which elicits the PDF
\begin{equation}
    \phi(x) = \frac{k}{\lambda} \left(\frac{x}{\lambda}\right)^{k-1} e^{-(x/\lambda)^k}
\end{equation}
with scale parameter $\lambda>0$, shape parameter $k>0$, and $x \geq 0$.
\end{mydefn}

First, a linear sum of independent but not identically distributed exponential random variables, as  the case for the polytopic target set constraint, results in a hypoexponential distribution. While a closed form expression of the cumulative distribution function exists, a closed form expression of the constraint would result in a reverse convex constraint. Further, as the cumulative distribution function is not invertible, quantile methods cannot be used. 

Second, the squared difference of exponential random variables, as is the case with the collision avoidance constraint, results in the sum of Weibull random variables. The PDF, CDF, and characteristic function of a sum of Weibull random variables can only be expressed as an infinite summation \cite{Yilmaz2009, Garcia2021}. Thus, closed form evaluations of the chance constraint probabilities are practically impossible. At present, methods that create bounds based on moments are the only methods that allow for almost surely satisfaction of each chance constraint.

\subsubsection{Experimental Setup}

For this experiment, we presume there are three deputies such that $N_v=3$. We presume the admissible control set is $\mathcal{U}_i = [-0.75, 0.75]^2 N\cdot \Delta$t${}^{-1}$ and time horizon $N=8$, corresponding to 8 minutes of operation. The performance objective is based on fuel consumption,
\begin{equation}
    J(\vec{U}_1, \dots, \vec{U}_{N_v}) = \sum^{N_v}_{i=1} \vec{U}_i^\top \vec{U}_i
\end{equation}

The terminal sets $\mathcal{T}_i(N)$ are $5\times 5$m boxes centered around desired terminal locations in $x,y$ coordinates with velocity bounded in both directions by $[-0.1, 0.1]$m/s. For collision avoidance, we presume that each deputy must remain at least $r=12$m away from each other, hence $S = \begin{bmatrix} I_{2} & 0_{2} \end{bmatrix}$ to extract the positions. Violation thresholds for terminal sets and collision avoidance are $\alpha = \gamma = 0.075$. The chance constraints are defined as
\begin{align}
    \pr{ \bigcap_{i=1}^3 \bvec{x}_i(N) \in \mathcal{T}_{i}(N) } &\geq 1-\alpha \label{eq:terminal}\\
    \pr{ \bigcap_{k=1}^{8} \bigcap_{i,j=1}^{3} \left\| S \left(\bvec{x}_i(k) -  \bvec{x}_j(k)\right)\right\| \geq r } &\geq 1-\gamma  \label{eq:avoidance_AV}
\end{align}
Figure \ref{fig:demo} provides a graphic representation of the demonstration presented.

As has been established \cite{ono2008iterative, Gorski2007}, biconvexity associated with having both risk allocation and control variables can be addressed in an iterative fashion, by alternately solving for the risk allocation variables, then for the control. However, for our demonstration, to isolate the impact of the one-sided Vysochanskij-Petunin inequality, we presume a fixed risk allocation. We uniformly allocate risk such that 
\begin{align}
    \pr{ \left\| S \left(\bvec{x}_i(k) -  \bvec{x}_j(k)\right)\right\| \geq r } &\geq 1-\hat{\gamma} & \forall \; i,j,k \label{eq:avoidance_AV_each}
\end{align}
where $\hat{\gamma} = \frac{\gamma}{24 \text{ constraints}} = \frac{.15}{24} = 3.125\times 10^{-3}$. These values remain constraint throughout the iterative solution finding process.

We define the solution convergence thresholds for the convex-concave procedure as both the difference of sequential performance objectives as less than $10^{-6}$ and the sum of slack variables as less than $10^{-8}$. Difference of convex programs were limited to 100 iterations. The first order approximations of the reverse convex constraints were initially computed assuming no system input.

For a random variable $\boldsymbol{x} \sim Exp(\lambda)$, where $\lambda$ is the rate parameter,
\begin{equation}\label{eq:Exponential_exp}
    \ex{\boldsymbol{x}^n} = \frac{n!}{\lambda^n} \qquad \forall \ n \in \N
\end{equation}
Hence, Assumption \ref{assm:moments} is valid. For brevity, the derivation of expectation, variance, and covariance terms for the collision avoidance constraint \eqref{eq:norm_reform_p5} can be found in Appendix \ref{appx:exp_deriv}. 

For the target set constraint, we can determine that the chance constraint is unimodal as the exponential distribution is a strongly unimodal distribution, as per Definition \ref{defn:strong_unimodal}. Hence, the affine constraint is unimodal. However, the Weibull random variables that result in the collision avoidance constraint are not strong unimodal. Here, unimodality of the constraint was validated numerically via Algorithm \ref{algo:unimodal} for each vehicle pair and each time step after computing the solution. Validation was completed with randomly sampled 50,000 disturbances. 

\subsubsection{Comparison Methodology}

We compare our method against the method in \cite{paulson2017stochastic}, the predecessor of the method proposed in this work based on Cantelli's inequality. This approach is effective for and has been demonstrated on systems which have target constraints and can be solved via convex optimization. We extend this method to accommodate 2-norm based collision constraints (as in Section \ref{ssec:collision_reform}) for the purpose of comparison with our own approach. We do not consider methods based on Chebyshev's inequality because they have shown to be less effective than \cite{paulson2017stochastic} in a target constraint problem \cite{Xu2019}.  

\begin{thm}[Cantelli's inequality \cite{Boucheron2013}]
Let $\boldsymbol{x}$ be a real valued random variable with finite expectation $\ex{\boldsymbol{x}}$ and finite, non-zero standard deviation $\std{\boldsymbol{x}}$. Then, for any $\lambda > 0$, 
\end{thm}
\begin{equation} \label{eq:cantelli}
    \pr{\boldsymbol{x} - \ex{\boldsymbol{x}}  \geq  \lambda \std{\boldsymbol{x}}} \leq \frac{1}{\lambda^2+1}
\end{equation}

\subsubsection{Experimental Results}

 \begin{figure}
    \centering
    \includegraphics[width=0.9\columnwidth]{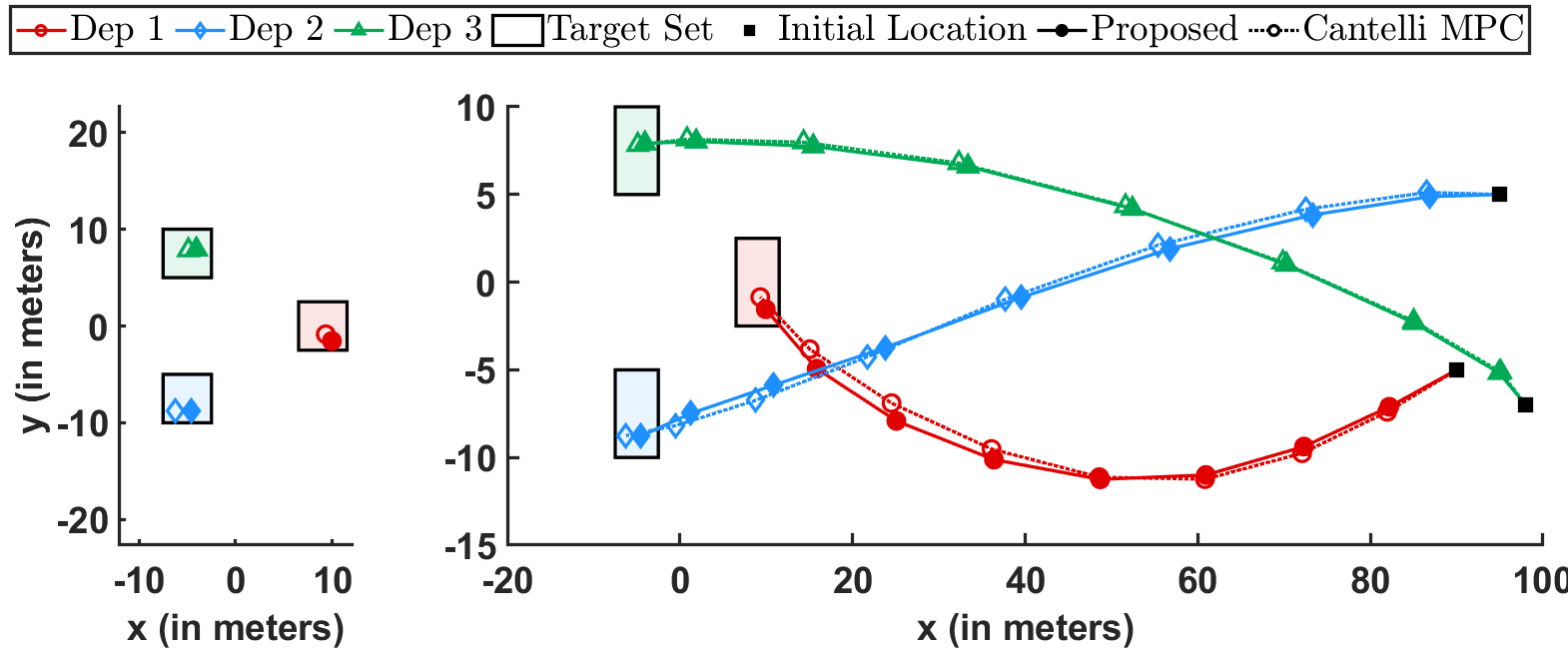}
    \caption{Comparison of mean trajectories between proposed method (solid line with filled in markers), and MPC with Cantelli's inequality \cite{paulson2017stochastic} (dotted line with white markers) for planar CWH dynamics with exponential disturbance. The full trajectory is displayed on the right and the terminal state is on the left. We see the two methods had similar trajectories but notice the proposed method was closer to the boundary of the target sets.}
    \label{fig:exp_traj_compare}
\end{figure}

\begin{table}
    \caption{Comparison of Solution and Computation Time for CWH Dynamics with Exponential Disturbance.}
    \centering
    \begin{tabular}{lcc}
         \hline \hline
         Metric &  Proposed Method & MPC with Cantelli's Inequality \cite{paulson2017stochastic}\\
         \hline 
         Solve Time & 6.5230 s  & 10.7012  s\\ 
         Iterations &  9  & 13 \\ 
         Solution Cost & $0.1156$ &   $0.1244$ \\ \hline
    \end{tabular}
    \label{tab:exp_stats}
\end{table}

\begin{table}
    \caption{Constraint Satisfaction for CWH Dynamics with Exponential Disturbance, with $10^4$ Samples and Probabilistic Violation Threshold of $\alpha = \gamma =0.075$.}
    \centering
    \begin{tabular}{lcc}
         \hline 
         \hline
         Constraint &  Proposed Method & MPC with Cantelli's Inequality \cite{paulson2017stochastic}  \\ \hline 
         \eqref{eq:terminal} & 0.9999   & 1.0000 \\ 
         \eqref{eq:avoidance_AV} & 1.0000 &  1.0000  \\
         \hline
    \end{tabular}
    \label{tab:exp_constraint}
\end{table}

The resulting trajectories are shown in Figure \ref{fig:exp_traj_compare}. We see that the two solutions result in similar trajectories. The most noticeable difference is that the trajectory of the proposed method was consistently closer to the boundary of the target set. The solution cost, iterations needed to converge, and computation times are shown in Table \ref{tab:exp_stats}. In all three categories, the proposed method performed better than the method of \cite{paulson2017stochastic}. To assess constraint satisfaction, we generated $10^4$ Monte Carlo sample disturbances for each approach. Table \ref{tab:exp_constraint} shows that while both methods were conservative, the proposed method was less conservative. 

As this example demonstrates, we can make probabilistic guarantees for disturbances that may arise in common circumstances. As discussed earlier, these distributional assumptions made in this example result in complicated distributions that lack analytical form. It is in distribution assumptions like these made in this example where this method will thrive.

\subsection{Gaussian Disturbance} \label{ssec:gauss_ex}

We include an example with a Gaussian disturbance to facilitate comparison with more conventional methods. In this example, we simplify the comparison example to only consider a convex joint chance constraint with a time-varying target set, as in Section \ref{ssec:target_reform}.

\subsubsection{Experimental Setup}

\begin{figure}
    \centering
    \includegraphics[width=0.6\columnwidth]{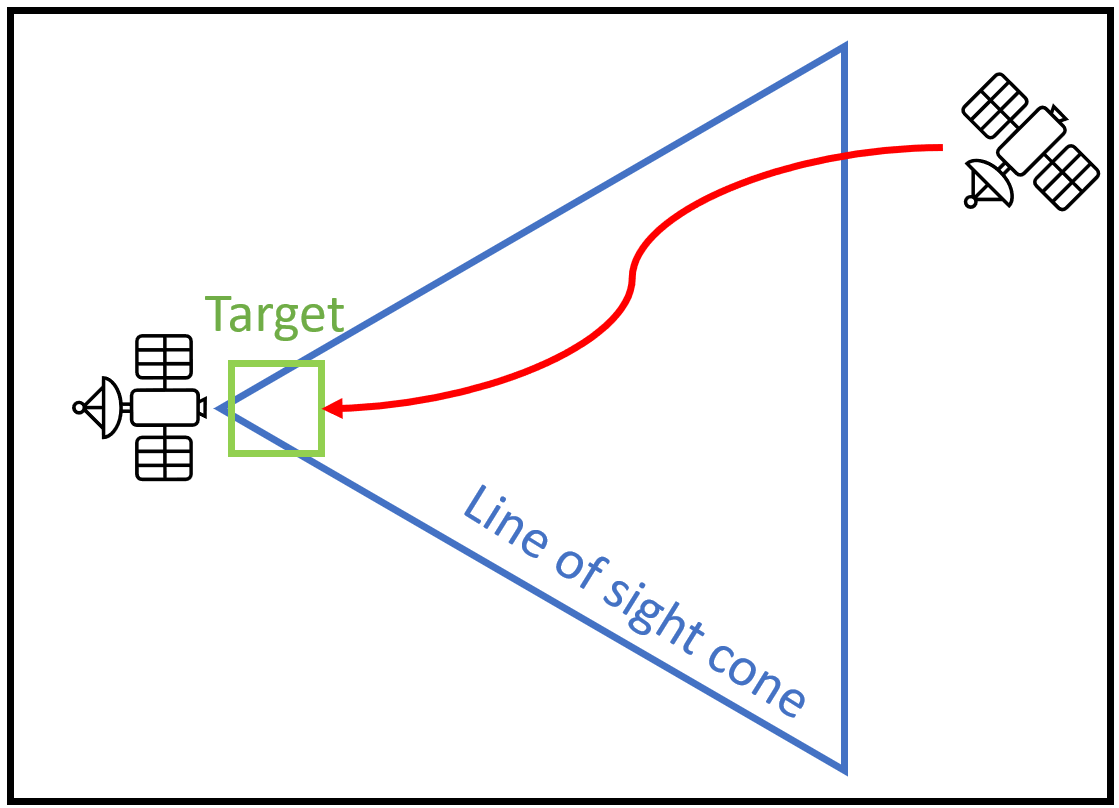}
    \caption{Graphic representation of the problem posed in Section \ref{ssec:gauss_ex}. Here, the dynamics of the deputy is perturbed by additive Gaussian noise. We attempt to find a control sequence that allows the deputy to rendezvous with the chief while meeting probabilistic time varying target set requirements.}
    \label{fig:gauss_demo}
\end{figure}

For this experiment, we presume there is a single deputy that must stay within a predefined line of sight cone and reach a terminal target set as shown in Figure \ref{fig:gauss_demo}. We presume the admissible control set is $\mathcal{U}_i = [-0.1, 0.1]^2 N\cdot \Delta$t${}^{-1}$ and time horizon $N=5$, corresponding to 5 minutes of operation. The performance objective is based on fuel consumption,
\begin{equation}
    J(\vec{U}_1) =  \vec{U}_1^\top \vec{U}_1
\end{equation}

The line-of-sight cone is defined by the inequalities
\begin{equation}
    \begin{split}
        -x + 2y \leq & \; 0 \\
        -x - 2y \leq & \; 0 \\
        x \leq & \; 10
    \end{split}
\end{equation}
The terminal sets $\mathcal{T}(N)$ is a $2\times 1$m near the origin with velocity bounded in both directions by $[-0.1, 0.1]$m/s. The violation thresholds for joint target set constraint is $\alpha = 0.05$. The chance constraint is defined as
\begin{equation}
    \pr{ \bigcap_{k=1}^5 \bvec{x}_1(k) \in \mathcal{T}_1(k) } \geq 1-\alpha \label{eq:terminal_gauss}\\
\end{equation}

We presume the disturbance is Gaussian,
\begin{equation}
    \bvec{w}_1(k) \sim N \left( \vec{0}, \mathrm{diag}\left( 10^{-3}, 10^{-3}, 10^{-8}, 10^{-8} \right) \right)
\end{equation}
Using the properties of the Gaussian disturbance, we know that all moments exist such that Assumption \ref{assm:moments} is valid. Further, affine summations of Gaussian disturbances are still Gaussian. Hence, each target set constraint is unimodal, validating Assumption \ref{assm:unimodal}.

\subsubsection{Comparison Methodologies}

Here, we compare the proposed methodology against a broader field of chance constrained stochastic optimal control methods. Several methods exist to solve convex chance constraints in a Gaussian regime. Hence, we select comparison methodologies that have been used extensively to solve chance constrained problems with Gaussian disturbances but can also handle non-Gaussian disturbances. Specifically, we compare the proposed method with quantile approach in \cite{blackmore2011chance, PrioreACC21}, the scenario approach in \cite{calafiore2006scenario, Campi2018TAC}, and the particle control approach in \cite{blackmore2010_particle}. 

The quantile approach results in a reformulation that is a convex in the input and the Gaussian quantile function. The quantile method allows for almost surely guarantees of chance constraint satisfaction as the disturbance is Gaussian. The particle control approach relies on sample disturbances and the chance constraint reformulation results in a mixed integer linear program. The particle control approach can only guarantee chance constraint satisfaction asymptotically as the number of samples goes to infinity. To minimize computational complexity, we select 200 sample disturbances to compute the optimal control trajectory with the particle control approach. 

Like the particle control approach, the scenario approach relies on samples to compute an optimal controller. The reformulation of the scenario approach results in a linear program. The scenario approach can guarantee chance constraints up to a probabilistic confidence bound $\delta$. By setting the confidence bound to a sufficiently small value, the probabilistic guarantees of the scenario approach closely resemble that of the proposed method. We compute the number of samples required for the scenario approach with the formula \cite{Campi2008}
\begin{equation}
    N_s \geq \frac{2}{\alpha}\left( \ln{\frac{1}{\delta}} + N_o \right)
\end{equation}
where $N_s$ is the number of samples required and $N_o$ is the number of optimization variables. Here, $N_o = 10$, and we choose $\delta = 10^{-16}$ and $N_s= 937$.

We expect the proposed method to result in more conservative solutions compared with these approaches. This stems from the conservative nature of the one-sided Vysochanskij-Petunin inequality \cite{Mercadier2021}. However, we also expect to see the proposed method compute solutions in less time than the comparison methods. We expect this as the proposed method doesn't rely on samples as the scenario and particle control method, and the simplicity of the proposed reformulation in comparison to the quantile approach.

\subsubsection{Experimental Results}

\begin{table*}
    \caption{Comparison of Computation Time, Solution Cost, and Constraint Satisfaction for CWH Dynamics with Multivariate Gaussian Disturbance with violation threshold $\alpha = 0.05$. Chance constraint satisfaction was measured as a ratio of $10^4$ samples satisfying the constraint.}
    \centering
    \begin{tabular}{lccccc}
         \hline \hline
         Metric             & Proposed Method       & Quantile Method \cite{blackmore2011chance, PrioreACC21} & Scenario Approach \cite{calafiore2006scenario, Campi2018TAC} & Particle Control \cite{blackmore2010_particle}
         \\ \hline
         Solve Time (sec)   & 0.1842                        & 1.0377                    & 3.8575                    & 30.5232 \\ 
         Solution Cost      & $6.0075 \!\times\! 10^{-4}$     & $5.1885 \!\times\! 10^{-4}$ & $5.3353 \!\times\! 10^{-4}$ & $5.1865 \!\times\! 10^{-4}$ \\ 
         Satisfaction of \eqref{eq:terminal_gauss} & 1.0000 & 0.9539                    & 0.9937                    & 0.9449 \\
         \hline
    \end{tabular}
    \label{tab:gauss_stats}
\end{table*}

\begin{figure}
    \centering
    \includegraphics[width=0.9\columnwidth]{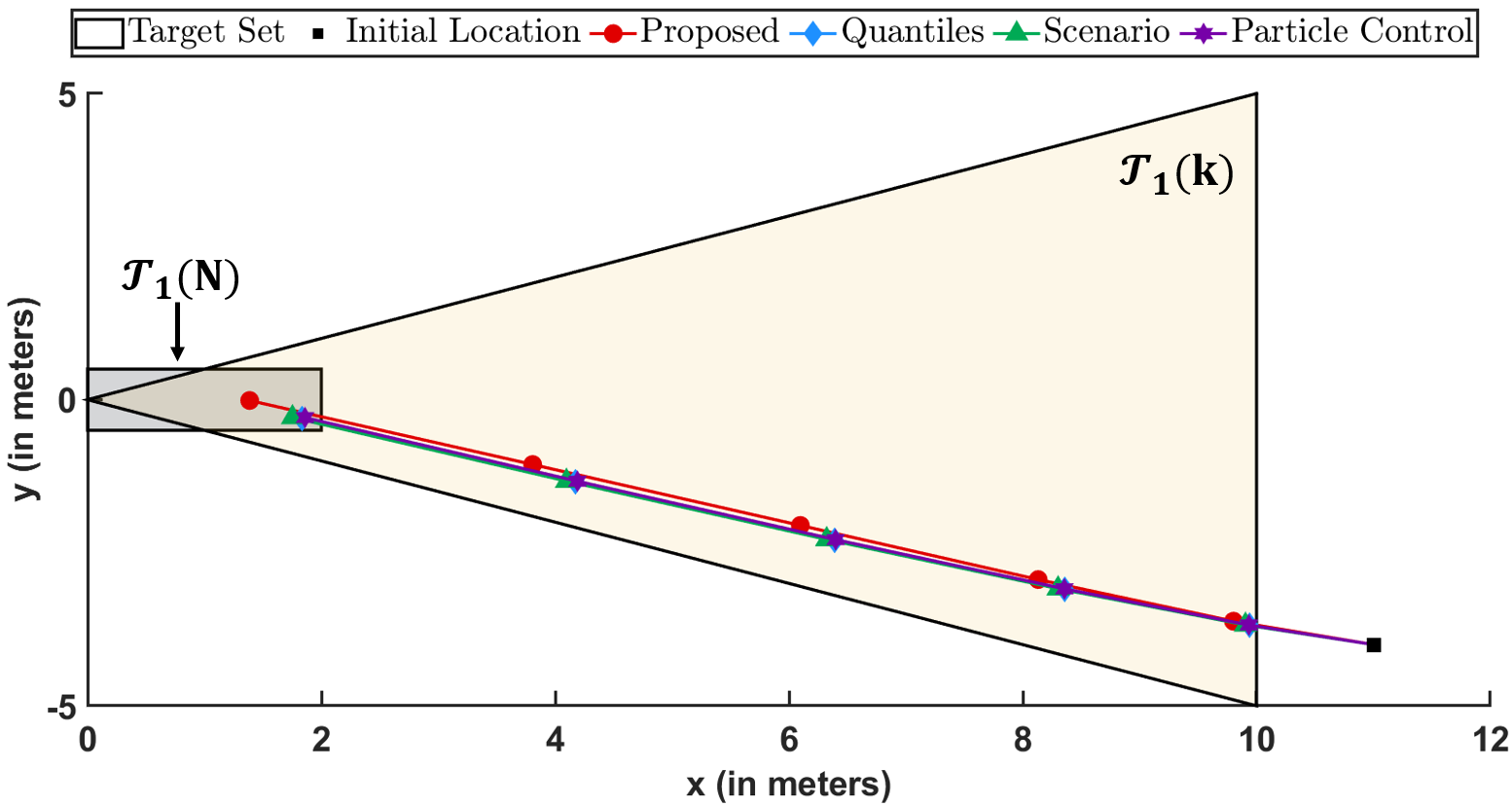}
    \caption{Comparison of mean trajectories between proposed method (red, circle), quantile-based approach \cite{blackmore2011chance, PrioreACC21} (blue, diamond), the scenario approach \cite{calafiore2006scenario, Campi2018TAC} (green, triangle), and the particle control approach \cite{blackmore2010_particle} (purple, 6-pointed star) for CWH dynamics with multivariate Gaussian disturbance. Here, we observe the trajectories are very similar. Note that the quantile approach and the particle control approach had nearly identical trajectories. This makes it difficult to see the trajectory of the quantile approach in this figure.}
    \label{fig:gauss_traj_compare}
\end{figure}

The resulting trajectories are not very different between the four methods as shown in Figure \ref{fig:gauss_traj_compare}. The most notable difference is that the trajectory of the proposed method is further from the boundary of the target sets, implying conservatism of the trajectory, as expected. This is also shown in Table \ref{tab:gauss_stats}. Here, we see the proposed method has higher chance constraint satisfaction and larger solution cost. We note that in empirical testing of chance constraint satisfaction, only particle control was not able to meet the required probability violation threshold as expected.

Table \ref{tab:gauss_stats} shows that the proposed method was able to compute the solution in significantly less time. Indeed, the proposed method was an order of magnitude faster than the quantile approach and the scenario approach, and two orders of magnitude faster than the particle control approach. 

As shown in this example, the method sacrifices optimality for broad applicability. In this particular case, the sacrifice was a solution cost that was approximately $13\%$ larger than the compared methods. However, the computational benefits, broad applicability of this method, and almost surely guarantees of chance constraint satisfaction present a strong case to use this method in instances where the improved speed is important. 

\section{Conclusions and Future Work} \label{sec:conclusion}

We proposed a framework to solve chance-constrained stochastic optimal control problems for LTI systems subject to arbitrary disturbances under moment and unimodality assumptions. This work focuses on probabilistic requirements for polytopic target sets and 2-norm based collision avoidance constraints. Our approach relies on the one-sided Vysochanskij–Petunin inequality to reformulate joint chance constraints into a series of inequalities that can be readily solved as a difference of convex functions optimization problem. We demonstrated our method on a multi-satellite rendezvous scenario under exponential and Gaussian disturbance assumptions and compare with an MPC approach using Cantelli's inequality (the predecessor of this work), a quantile-based approach, the scenario approach, and the particle control approach. We showed that this approach is amenable to disturbances that prove challenging or impossible to solve with other methods and demonstrated the proposed method has computational benefits in comparison to other commonly used methods.

Methodologically, we are interested in exploring probabilistic inequalities that result in less conservative bounds. For scenarios in which the disturbance in unknown and samples are available, we are currently exploring moment-based approaches that rely on sample approximations of moments and provide probabilistic guarantees.

\appendix

\subsection{Numerical Evaluation of Unimodality} \label{appx:unimodal}

Algorithm \ref{algo:unimodal} is constructing an affine approximation of empirical cumulative distribution function then testing whether there is a single inflection point by comparing the slopes of the affine segments.

\begin{algorithm}
    \DontPrintSemicolon
    \caption{Numerical check for unimodality.}
	\label{algo:unimodal}
	\textbf{Input}: Empirical cumulative distribution function points $(x_i, \hat{F}(x_i))$ for samples $x_i$ with $i \in \Nt{1}{N_s}$, and maximum error threshold $\xi$.\;
	\textbf{Output}: 1 if unimodal or 0 if not unimodal \;
	i $\gets 0$ \;
	$\mathbb{S} \gets \varnothing$ \;
    \While{$i < N_s$}{
    	\For{$j = N_s$ \textbf{to} $i+1$ \textbf{by} $-1$}{
        	$\underline{m} \gets \frac{\hat{F}(x_j)-\hat{F}(x_i)}{x_j-x_i}$\; 
        	$\underline{b} \gets \hat{F}(x_j) - x_j \times \underline{m}$\; 
        	\For{$k=i+1$ \textbf{to} $j-1$ \textbf{by} $1$}{
            	$\epsilon_k = \hat{F}(x_k) - (p_y \times \underline{m} + \underline{b})$ \;
            	\If{$\epsilon_k > \xi$}{\textbf{next} $j$}
            }	
            $\mathbb{S} \gets \mathbb{S} \cup \{\underline{m}\}$\;
            \textbf{break}\;
    	}
    	$i \gets j$ \;
	}
	$w \gets 0$\; 
	$N_c \gets card(\mathbb{S})$ \qquad\qquad \# cardinality of set $\mathbb{S}$\;
	\For{$i = 2$ \textbf{to} $N_c$ \textbf{by} $1$}{
	    \# $\mathbb{S}_i$ is the $i^{th}$ element of $\mathbb{S}$\;
	    \eIf{$\mathbb{S}_{i} \geq \mathbb{S}_{i-1}$}{
	        \If{$w = 1$}{
	            \textbf{return} 0
	        }
	    }{
	        $w \gets 1$
	    }
	}
	\textbf{return} 1\;
\end{algorithm}

\subsection{Derivation of Norm Expectation and Variance in Exponential Case} \label{appx:exp_deriv}

We keep with the notation used in Section \ref{ssec:collision_reform}. Since we assumed the disturbances are independent and identically distributed, from \eqref{eq:Exponential_exp} we find
\begin{equation}
\begin{split}\label{eq:weibul_norm_zero}
    \ex{\bvec{z}} & = 0_{q\times 1} \\
    \var{\bvec{z}} & = 2 S \mathcal{D}(k) \var{\bvec{W}_i} \mathcal{D}^{\top}(k) S^{\top} 
\end{split}
\end{equation}
where 
\begin{equation}
\begin{split}
    & \var{\bvec{W}_i} \\
    & \ = \mathrm{diag} \left(20^{-2} \cdot I_2, 10^{-8} \cdot I_2, \dots, 20^{-2} \cdot I_2, 10^{-8} \cdot I_2 \right)
\end{split}
\end{equation}
Next, from \eqref{eq:weibul_norm_zero}
\begin{equation}
        \ex{\bvec{z}^{\top} \bvec{z}}  = \tr{\var{\bvec{z}}}
\end{equation}
Next, we find $\var{\bvec{z}^{\top} \bvec{z}}$. For brevity, we denote $\bvec{W}_i - \bvec{W}_j$ as $\bvec{\mathcal{W}}$. Then,
\begin{subequations}
\begin{align}
    & \var{\bvec{z}^{\top} \bvec{z}} \nonumber \\
    & \ = \var{\bvec{\mathcal{W}}^{\top} \mathcal{D}^{\top}(k) S^{\top} S  \mathcal{D}(k)\bvec{\mathcal{W}}}\\
    & \ = \var{\sum_{p=1}^{Nn}\sum_{q=1}^{Nn} a_{pq}\bvec{\mathcal{W}}_{p} \bvec{\mathcal{W}}_{q}} \\
    & \ = \sum_{p=1}^{Nn}\sum_{q=1}^{Nn}  \sum_{r=1}^{Nn}\sum_{s=1}^{Nn} \cov{ a_{pq}\boldsymbol{\mathcal{W}}_{p} \boldsymbol{\mathcal{W}}_{q}}{ a_{rs}\boldsymbol{\mathcal{W}}_{r} \boldsymbol{\mathcal{W}}_{s}} 
\end{align}
\end{subequations}
where $a_{pq}$ is the $(p,q)$\textsuperscript{th} element of $\mathcal{D}^{\top}(k) S^{\top} S \mathcal{D}(k)$. Then
\begin{equation}
\begin{split}
    \sum_{p=1}^{Nn}&\sum_{q=1}^{Nn} \sum_{r=1}^{Nn}\sum_{s=1}^{Nn} \cov{ a_{pq}\bvec{\mathcal{W}}_{p} \bvec{\mathcal{W}}_{q}}{ a_{rs}\bvec{\mathcal{W}}_{r} \bvec{\mathcal{W}}_{s}} \\
    = & \; \sum_{p=1}^{Nn}\var{ a_{pp}\bvec{\mathcal{W}}_{p}^2}  + 4\sum_{1\leq p < q \leq Nn} \var{ a_{pq}\bvec{\mathcal{W}}_{p} \bvec{\mathcal{W}}_{q}} 
\end{split}
\end{equation}
Here, all remaining covariance terms is zero as each element is mutually independent by Assumptions \ref{assm:vec_ind} and \ref{assm:elem_ind}, and the first and third moments being zero. So,
\begin{subequations}
\begin{align}
    & \sum_{p=1}^{Nn} \var{ a_{pp}\bvec{\mathcal{W}}_{p}^2}  + 4 \sum_{1\leq p < q \leq Nn} \var{ a_{pq}\bvec{\mathcal{W}}_{p} \bvec{\mathcal{W}}_{q}} \nonumber \\
    & \ = \sum_{p=1}^{Nn} a_{pp}^2 \left (\ex{   \bvec{\mathcal{W}}_{p}^4} - \ex{   \bvec{\mathcal{W}}_{p}^2}^2 \right)  \\
    & \ \ + 4 \sum_{1\leq p < q \leq Nn} a_{pq}^2 \ex{\bvec{\mathcal{W}}_{p}^2} \ex{ \bvec{\mathcal{W}}_{q}^2} \nonumber\\
    & \ = 3 \sum_{p=1}^{Nn} a_{pp}^2 \ex{   \bvec{\mathcal{W}}_{p}^2}^2 \!+\! 2 \sum_{p=1}^{Nn}\sum_{q=1}^{Nn} a_{pq}^2 \ex{\bvec{\mathcal{W}}_{p}^2} \ex{ \bvec{\mathcal{W}}_{q}^2} 
\end{align}
\end{subequations}
as in this example $\ex{\bvec{\mathcal{W}}_{p}^4} = 6 \ex{ \bvec{\mathcal{W}}_{p}^2}^2$. Then, let $\vec{a}$ be a vector consisting of the diagonal elements of $\mathcal{D}^{\top}(k) S^{\top} S \mathcal{D}(k)$. So, 
\begin{subequations}
\begin{align}
    &  3 \sum_{p=1}^{Nn} a_{pp}^2 \ex{   \bvec{\mathcal{W}}_{p}^2}^2 + 2 \sum_{p=1}^{Nn}\sum_{q=1}^{Nn} a_{pq}^2 \ex{\bvec{\mathcal{W}}_{p}^2} \ex{ \bvec{\mathcal{W}}_{q}^2} \\
    & \ = 12 \vec{a}^{\top}\var{\bvec{W}_i}^2 \vec{a} \\
    & \quad + 8 \tr{\left( \mathcal{D}^\top(k) S^\top \var{\bvec{W}_i} S \mathcal{D}(k) \right)^2}\nonumber
\end{align}
\end{subequations}
Finally, we find $\cov{\bvec{z}}{\bvec{z}^{\top}\bvec{z}}$.
\begin{equation}
    \cov{\bvec{z}}{\bvec{z}^{\top}\bvec{z}} = \ex{\bvec{z} \bvec{z}^{\top}\bvec{z}} - \ex{\bvec{z}} \ex{ \bvec{z}^{\top}\bvec{z}}
\end{equation}
The second term is zero by \eqref{eq:weibul_norm_zero}. For a random vector, the expectation is a vector of the expectations of each element. Then for the $i$\textsuperscript{th} element,
\begin{equation} \label{eq:exp_norm_var_deriv_cov}
    \ex{\bvec{z}_{i} \bvec{z}^{\top}\bvec{z}} =  \ex{\bvec{z}_{i}^3} + \sum_{\substack{j=1\\j \neq i}}^{Nn}\ex{\bvec{z}_i}\ex{\bvec{z}_j^2}
\end{equation}
Since the first and third moments of $\bvec{z}$ are zero, then the sum is zero. Thus, $\cov{\bvec{z}}{\bvec{z}^{\top}\bvec{z}} = 0$.

\bibliography{main}

\begin{thebibliography}{10}

\bibitem{Idan2019}
M.~Idan and J.~L. Speyer, ``Characteristic function approach to smoothing of
  linear scalar systems with additive cauchy noises,'' in {\em 2019 27th
  Mediterranean Conference on Control and Automation (MED)}, pp.~238--243,
  2019.

\bibitem{vinod2019piecewise}
A.~P. Vinod, V.~Sivaramakrishnan, and M.~Oishi, ``Piecewise-affine
  approximation-based stochastic optimal control with gaussian joint chance
  constraints,'' in {\em Proc. Amer. Ctrl. Conf.}, pp.~2942--2949, 2019.

\bibitem{Sivaramakrishnan2021TAC}
V.~Sivaramakrishnan, A.~P. Vinod, and M.~Oishi, ``Convexified open-loop
  stochastic optimal control for linear non-gaussian systems,'' {\em
  arXiv:2010.02101}, 2021.

\bibitem{ono2008iterative}
M.~Ono and B.~Williams, ``Iterative risk allocation: A new approach to robust
  model predictive control with a joint chance constraint,'' in {\em IEEE Conf.
  Dec. \& Control}, pp.~3427--3432, 2008.

\bibitem{calafiore2006scenario}
G.~Calafiore and M.~Campi, ``The scenario approach to robust control design,''
  {\em {IEEE} Trans. Autom. Control}, vol.~51, no.~5, pp.~742--753, 2006.

\bibitem{campi2011sampling}
M.~Campi and S.~Garatti, ``A sampling-and-discarding approach to
  chance-constrained optimization: Feasibility and optimality,'' {\em J. Optim
  Theory Appl.}, vol.~148, no.~2, pp.~257--280, 2011.

\bibitem{care2014fast}
A.~Car{\`e}, S.~Garatti, and M.~C. Campi, ``Fast--fast algorithm for the
  scenario technique,'' {\em Ops. Res.}, vol.~62, no.~3, pp.~662--671, 2014.

\bibitem{Campi2018TAC}
M.~C. Campi, S.~Garatti, and F.~A. Ramponi, ``A general scenario theory for
  nonconvex optimization and decision making,'' {\em IEEE Trans. Autom.
  Control}, vol.~63, no.~12, pp.~4067--4078, 2018.

\bibitem{Boucheron2013}
S.~Boucheron, G.~Lugosi, and P.~Massart, {\em {Concentration Inequalities: A
  Nonasymptotic Theory of Independence}}.
\newblock Oxford University Press, 02 2013.

\bibitem{Zhou2013}
Z.~Zhou and R.~Cogill, ``Reliable approximations of probability-constrained
  stochastic linear-quadratic control,'' {\em Automatica}, vol.~49, no.~8,
  pp.~2435--2439, 2013.

\bibitem{Xu2019}
J.~Xu, T.~van~den Boom, and B.~De~Schutter, ``Model predictive control for
  stochastic max-plus linear systems with chance constraints,'' {\em IEEE
  Trans. Autom. Control}, vol.~64, no.~1, pp.~337--342, 2019.

\bibitem{Farina2015}
M.~Farina, L.~Giulioni, L.~Magni, and R.~Scattolini, ``An approach to
  output-feedback mpc of stochastic linear discrete-time systems,'' {\em
  Automatica}, vol.~55, pp.~140--149, 2015.

\bibitem{paulson2017stochastic}
J.~Paulson, E.~Buehler, R.~Braatz, and A.~Mesbah, ``Stochastic model predictive
  control with joint chance constraints,'' {\em Int'l J. Ctrl.}, pp.~1--14,
  2017.

\bibitem{Mercadier2021}
M.~Mercadier and F.~Strobel, ``A one-sided vysochanskii-petunin inequality with
  financial applications,'' {\em European Journal of Operational Research},
  vol.~295, no.~1, pp.~374--377, 2021.

\bibitem{casella2002}
G.~Casella and R.~Berger, {\em Statistical Inference}.
\newblock Duxbury advanced series in statistics and decision sciences, Cengage
  Learning, 2002.

\bibitem{Bertin1997}
E.~M.~J. Bertin, I.~Cuculescu, and R.~Theodorescu, {\em Strong unimodality},
  pp.~183--200.
\newblock Dordrecht: Springer Netherlands, 1997.

\bibitem{Ibragimov1956}
I.~A. Ibragimov, ``On the composition of unimodal distributions,'' {\em Theory
  of Probability \& Its Applications}, vol.~1, no.~2, pp.~255--260, 1956.

\bibitem{Gorski2007}
J.~Gorski, F.~Pfeuffer, and K.~Klamroth, ``Biconvex sets and optimization with
  biconvex functions: a survey and extensions,'' {\em Mathematical Methods of
  Operations Research}, vol.~66, pp.~373--407, Dec 2007.

\bibitem{boyd_convex}
S.~Boyd and L.~Vandenberghe, {\em Convex Optimization}.
\newblock Cambridge University Press, 2004.

\bibitem{boyd_dc_2016}
T.~{Lipp} and S.~{Boyd}, ``Variations and extension of the convex–concave
  procedure,'' {\em Optimization and Eng.}, vol.~17, pp.~263--287, 2016.

\bibitem{horst2000}
R.~Horst, P.~M. Pardalos, and N.~V. Thoai, {\em Introduction to global
  optimization}.
\newblock Springer Science \& Business Media, 2000.

\bibitem{cvx}
M.~Grant and S.~Boyd, ``{CVX}: Matlab software for disciplined convex
  programming, version 2.1.'' \url{http://cvxr.com/cvx}, Mar. 2014.

\bibitem{gurobi}
L.~Gurobi~Optimization, ``Gurobi optimizer reference manual,'' 2020.

\bibitem{MPT3}
M.~Herceg, M.~Kvasnica, C.~Jones, and M.~Morari, ``{Multi-Parametric Toolbox
  3.0},'' in {\em Proc. Euro. Ctrl. Conf.}, (Z\"urich, Switzerland),
  pp.~502--510, July 17--19 2013.

\bibitem{wiesel1989_spaceflight}
W.~Wiesel, {\em Spaceflight Dynamics}.
\newblock New York: McGraw--Hill, 1989.

\bibitem{Yilmaz2009}
F.~Yilmaz and M.-S. Alouini, ``Sum of weibull variates and performance of
  diversity systems,'' in {\em Proceedings of the 2009 International Conference
  on Wireless Communications and Mobile Computing: Connecting the World
  Wirelessly}, Iwcmc '09, (New York, NY, USA), p.~247–252, Association for
  Computing Machinery, 2009.

\bibitem{Garcia2021}
F.~D.~A. Garcia, F.~R.~A. Parente, G.~Fraidenraich, and J.~C. S.~S. Filho,
  ``Light exact expressions for the sum of weibull random variables,'' {\em
  IEEE Wireless Communications Letters}, vol.~10, no.~11, pp.~2445--2449, 2021.

\bibitem{blackmore2011chance}
L.~Blackmore, M.~Ono, and B.~Williams, ``Chance-constrained optimal path
  planning with obstacles,'' {\em {IEEE} Trans. Robot.}, vol.~27, no.~6,
  pp.~1080--1094, 2011.

\bibitem{PrioreACC21}
S.~Priore, A.~Vinod, V.~Sivaramakrishnan, C.~Petersen, and M.~Oishi,
  ``Stochastic multi-satellite maneuvering with constraints in an elliptical
  orbit,'' in {\em 2021 American Control Conference (ACC)}, pp.~4261--4268,
  2021.

\bibitem{blackmore2010_particle}
L.~Blackmore, M.~Ono, A.~Bektassov, and B.~C. Williams, ``A probabilistic
  particle-control approximation of chance-constrained stochastic predictive
  control,'' {\em IEEE Trans. on Robotics}, vol.~26, pp.~502--517, June 2010.

\bibitem{Campi2008}
M.~C. Campi, S.~Garatti, and M.~Prandini, ``The scenario approach for systems
  and control design,'' {\em IFAC Proceedings Volumes}, vol.~41, no.~2,
  pp.~381--389, 2008.
\newblock 17th IFAC World Congress.

\end{thebibliography}
\bibliographystyle{ieeetr}

\begin{IEEEbiography}[{\includegraphics[width=1in,height=1.25in,clip,keepaspectratio]{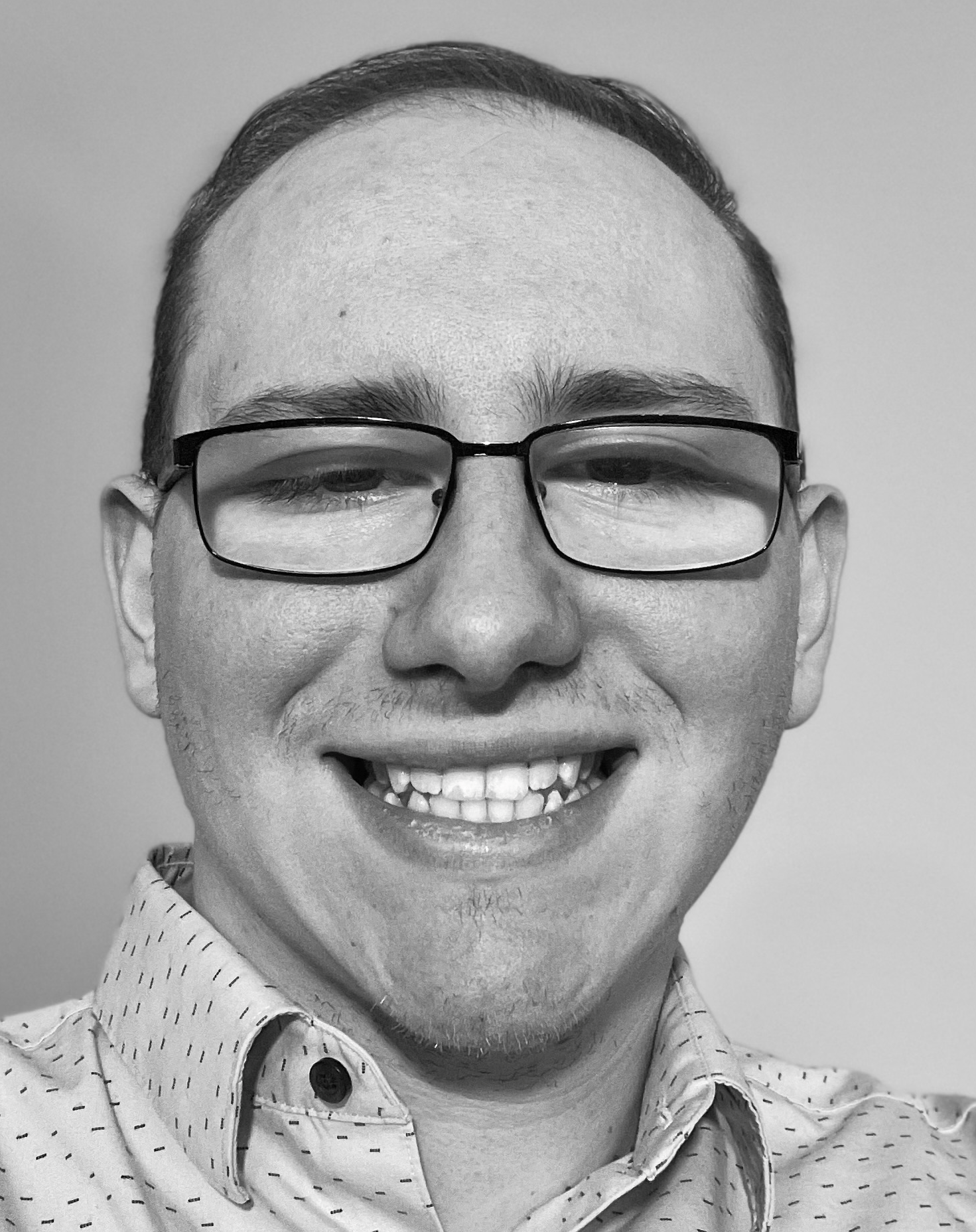}}]{Shawn Priore} (Student Member, IEEE) received the B.S. degree in economics and the M.B.A. degree from John Carroll University, University Heights, OH, USA, in 2016 and 2017, respectively, the M.S. degree in statistics from the University of Connecticut, Storrs, CT, USA. He is currently pursuing the Ph.D. degree in electrical engineering from the University of New Mexico, Albuquerque, NM, USA.

His research interests are in the area of chance constrained stochastic optimal control, autonomous systems, and probabilistic safety, with an emphasis on non-Gaussian disturbances. 

Mr. Priore is the recipient of the Department of Defense SMART Scholarship.
\end{IEEEbiography}

\begin{IEEEbiography}[{\includegraphics[width=1in,height=1.25in,clip,keepaspectratio]{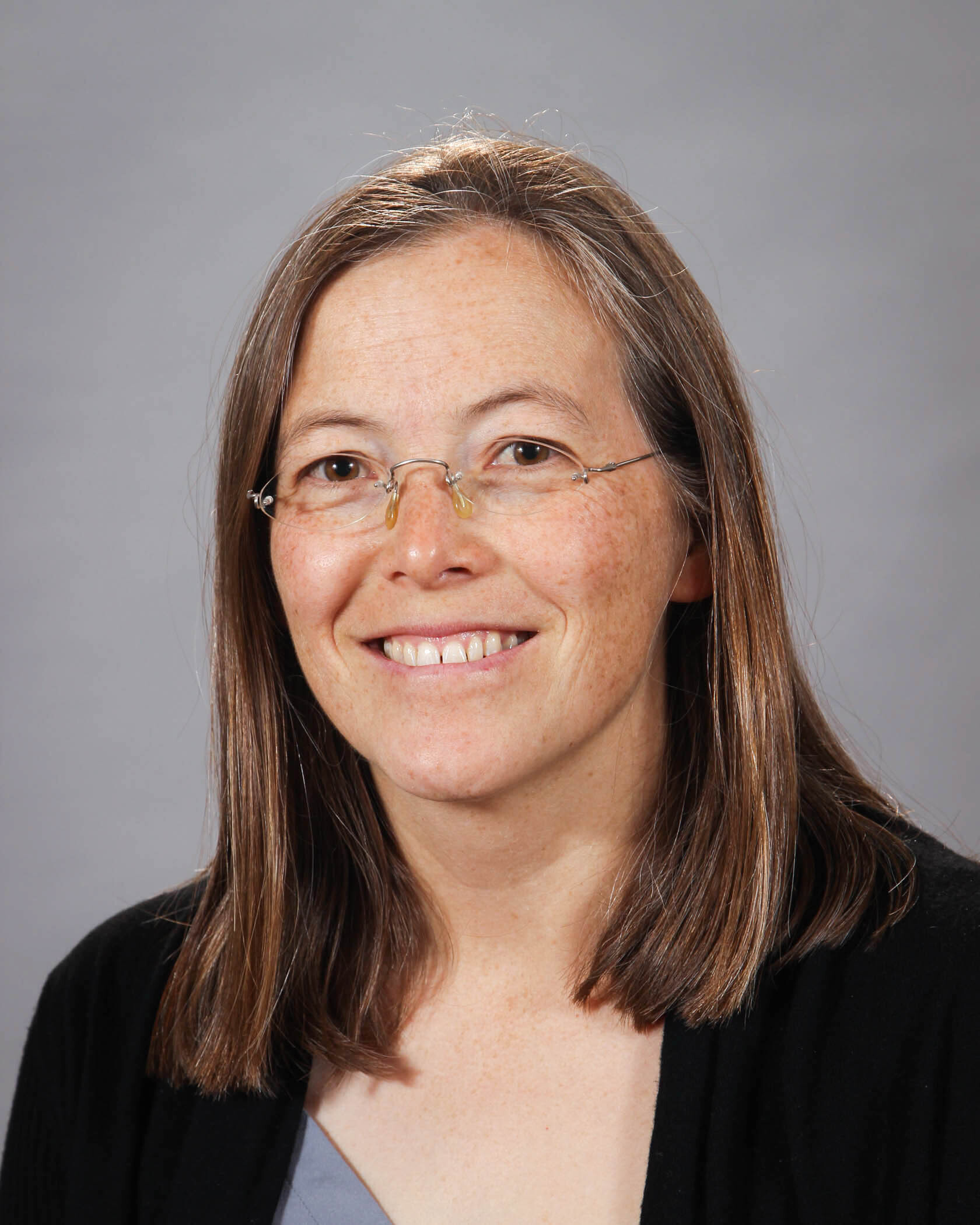}}]{Meeko Oishi} (SM '19, M '04, S '00) received the Ph.D. (2004) and M.S. (2000) in Mechanical Engineering from Stanford University (Ph.D. minor, Electrical Engineering), and a B.S.E. in Mechanical Engineering from Princeton University (1998).  

She is a Professor of Electrical and Computer Engineering at the University of New Mexico.  Her research interests include human-in-the-loop control, stochastic optimal control, and autonomous systems.  She previously held a faculty position at the University of British Columbia at Vancouver, and postdoctoral positions at Sandia National Laboratories and at the National Ecological Observatory Network.  

She is the recipient of the UNM Regents' Lectureship, the NSF CAREER Award, the UNM Teaching Fellowship, the Peter Wall Institute Early Career Scholar Award, the Truman Postdoctoral Fellowship in National Security Science and Engineering, and the George Bienkowski Memorial Prize, Princeton University. She was a Visiting Researcher at AFRL Space Vehicles Directorate, and a Science and Technology Policy Fellow at The National Academies.
\end{IEEEbiography}

\end{document}